\RequirePackage{ifpdf}
\documentclass{PoS}

\usepackage{breakurl} 
\usepackage{booktabs} 
\usepackage{multirow} 
\usepackage{cancel}  
\usepackage{listings}
\usepackage{verbatim}
\usepackage{alltt} 
\usepackage{graphicx}
\usepackage{amsmath}
\usepackage{amssymb}
\usepackage[mathscr]{eucal}
\usepackage{fancyvrb} 
\usepackage[hyphens]{url} 
\usepackage{color}     
\let\normalcolor\relax  

\def\mathswitch#1{\relax\ifmmode#1\else$#1$\fi}
\def\mathswitchr#1{\relax\ifmmode{\mathrm{#1}}\else$\mathrm{#1}$\fi}
\newcommand{\PW}{\mathswitchr W}
\newcommand{\PZ}{\mathswitchr Z}

\newcommand{\Pe}{\mathswitchr e}
\newcommand{\Pb}{\mathswitchr b}
\newcommand{\Pt}{\mathswitchr t}

\newcommand{\Pf}{\mathswitchr f}
\newcommand{\MW}{\mathswitch {M_\PW}}
\newcommand{\MZ}{\mathswitch {M_\PZ}}
\newcommand{\GZ}{\mathswitch {\Gamma_\PZ}}

\newcommand{\as}{\alpha_{\mathrm s}}
\newcommand{\at}{\alpha_\Pt}

\newcommand{\seff}[1]{\sin^2\theta_{\rm eff}^{\rm #1}}

\newcommand{\re}{\Re e \,}

\newcommand{\OO}{{\mathcal O}}

\newcommand{\sss}[1]{\scriptscriptstyle{#1}}
\newcommand{\zb}{Z}

\newcommand{\sman}{s}
\newcommand{\tman}{t}

\newcommand{\lpar}{\left(}                            
\newcommand{\rpar}{\right)}

\newcommand{\ib  }{i}
\newcommand{\qf  }{Q_f  }

\newcommand{\qe  }{Q_e  }

\newcommand{\gadu}[1]{\gamma_{#1}}

\newcommand{\tcie}{I^{(3)}_e}
\newcommand{\tcif}{I^{(3)}_f}

\def\gfd{\gamma_5}

    \newcommand{\stws}{s^2_{\sss{W}}}
    \newcommand{\stwf}{s^4_{\sss{W}}}

\title{%
{\flushright{ 
\small \texttt{HU-EP 16/36}
\\
\small \texttt{KW 16-003}
\\[8mm]
}}
30 years, some 700 integrals, and 1 dessert
\\
{~~~~~~~~~~~~~~~~~~~~~~~~~~~~~~~{\small or:}}
\\
Electroweak two-loop corrections to the $Z{\bar b}b$ vertex
\\
}

\ShortTitle{30 years, some 700 integrals, and 1 dessert
}

\author{I. Dubovyk$^{a}$, 
A. Freitas$^{b}$,
{J. Gluza}$^{c}$,
\speaker{T. Riemann}$^{,c,d}$, 
J. Usovitsch$^e$
\\
\\
\llap{$^{a}$} II. Institut f{\"u}r Theoretische Physik, Universit{\"a}t Hamburg,
22761 Hamburg,  Germany 
\\
\llap{$^{b}$}
Pittsburgh Particle physics, Astrophysics \& Cosmology Center
(PITT PACC), Department of Physics \& Astronomy, University of Pittsburgh,
Pittsburgh, PA 15260, USA 
\\
\llap{$^c$} Institute of Physics, University of Silesia, Uniwersytecka 4, PL-40007 Katowice, Poland 
\\
\llap{$^d$} 15711 K{\"o}nigs Wusterhausen, Germany
\\
\llap{$^e$} 
Institut f{\"u}r Physik, Humboldt-Universit{\"a}t zu Berlin, 12489 Berlin, Germany

\vspace*{3mm}
E-mails: 
\email{e.a.dubovyk@gmail.com}, \email{afreitas@pitt.edu}, \email{janusz.gluza@us.edu.pl}, \email{tordriemann@gmail.com},
\email{jusovitsch@googlemail.com}
}

\abstract{\hspace*{0cm}
The one-loop corrections to the weak mixing angle 
$\seff{b}$, derived from the $Z{\bar b}b$ vertex, are known since 1985.
It took another 30 years to calculate the complete electroweak two-loop corrections to $\seff{b}$.
The main obstacle was the calculation of the ${\cal O}(700)$ bosonic two-loop vertex integrals with up to three mass scales, at $s=M_Z^2$.
We did not perform the usual integral reduction and master evaluation, but chose a completely numerical approach, using two different 
calculational chains.
One method relies on publicly available sector decomposition implementations.
Further, we derived Mellin-Barnes (MB) representations, exploring the publicly available MB suite. We had to supplement the MB suite by 
two new packages: AMBRE~3, a Mathematica program, for the efficient treatment of non-planar integrals and MBnumerics for advanced numerics 
in the Minkowskian space-time. 
Our preliminary result for LL2016, the ``dessert'', for the electroweak bosonic two-loop contributions to $\seff{b}$ is:
\\
~~~~ $\Delta \seff{b(\alpha^2,\rm bos)} = \sin^2\theta_W  ~ \Delta\kappa_\Pb^{(\alpha^2,\rm bos)}$, with
$\Delta\kappa_\Pb^{(\alpha^2,\rm bos)} = -1.0276 \times 10^{-4}$.
\\
This contribution is about a quarter of the corresponding fermionic corrections and of about the same magnitude as several of the 
known higher-order QCD corrections.
The $\seff{b}$ is now predicited in the Standard Model with 
a relative error of $10^{-4}$ \cite{Dubovyk:2016aqv}.
}

\FullConference{Loops and Legs in Quantum Field Theory - LL 2016, \\
                24 - 29 April 2016            \\
                Leipzig, Germany}

\begin{document}\allowdisplaybreaks
\begin{center}
 \includegraphics[width=.9\linewidth]{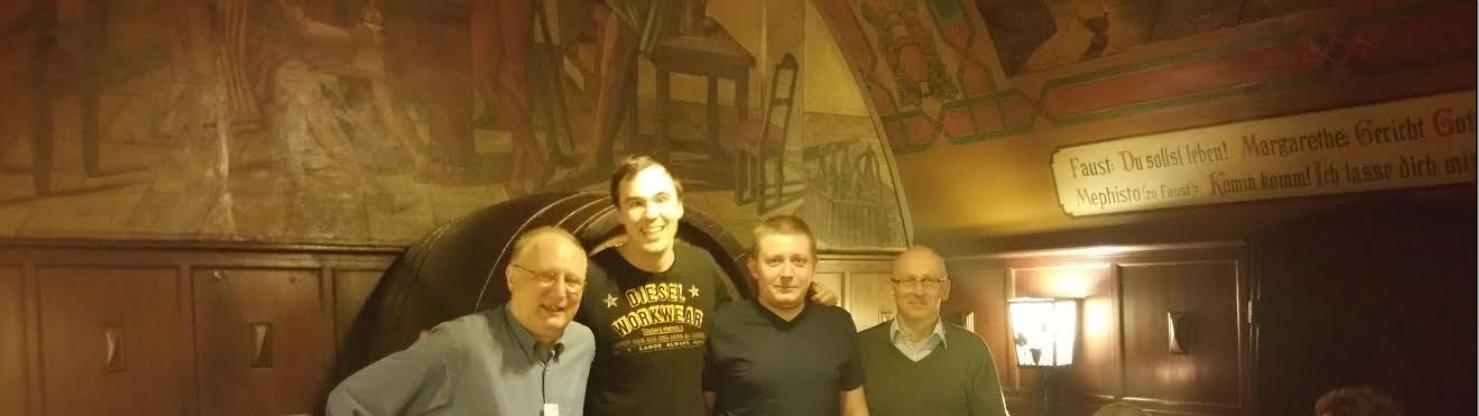}
\\
{\footnotesize \it \hspace*{0.2cm}
In Auerbachs Keller, where Faust met Mephisto: T. Riemann, J. Usovitsch, I. Dubovyk, J. Gluza}
\end{center}

\section*{Preface} 


 {\it 
\begin{quote}
Der Wahnsinn
\\
ist nur eine schmale Br{\"u}cke
\\
die Ufer sind Vernunft und Trieb

{\footnotesize Rammstein, \href{http://www.magistrix.de/lyrics/Rammstein/Du-Riechst-So-Gut-26274.html}{
http://www.magistrix.de/lyrics/Rammstein/Du-Riechst-So-Gut-26274.html} 
 \footnote{
 {\it Insanity -- is just a narrow bridge -- the shores are reason and urge},
 Rammstein, 
\href{http://lyricstranslate.com/de/du-riechst-so-gut-you-smell-so-good.html}{
http://lyricstranslate.com/de/du-riechst-so-gut-you-smell-so-good.html} (22.8.2016)
 }
}
\end{quote}}

LL2016 in Leipzig was the 13$^{th}$ edition, and it was my last Loops and Legs conference as an organizer, and perhaps also as a participant.
I founded it, together with Johannes Bl{\"u}mlein and Martina Mende, in April 1992 as a bi-annual event; it was a follow-up of the 1989 
November changes in Germany. 
The aim was an overview of the recent developments in perturbative quantum field theory with focus on applications 
to precision collider experiments.
We attempted a mix of all the relevant research directions, but also a mix of both younger and more experienced collegues.
From time to time we had to invent some modernizations like parallel sessions or online proceedings.
In the early years the focus was more at direct phenomenological applications, and now it is more on advanced technical developments.
The steady high scientific level has been guaranteed by the participants, and the discussions were top-level, lively and sometimes even 
hot.

My own research reflects the above observations.
Since 1977, I am engaged in complete electroweak radiative corrections for collider physics. 
A statement by a head of an institute: ``For me you are one who calculates integrals.''
My most successful project is ZFITTER (with Dima Bardin et al., see 
\href{http://sanc.jinr.ru/users/zfitter}{http://sanc.jinr.ru/users/zfitter}). 
ZFITTER became the standard software for the 
study of the Z boson resonance at LEP and elsewhere, and 
it was used to predict the masses of the top quark and the Higgs boson prior to their 
discoveries. 
Many PhD students used it.
Recently, ZFITTER even got illegally copied by non-experts in order to promote their carriers.     
In 1985, we calculated the one-loop corrections to $Z \to {\bar b}b$.
The corresponding FORTRAN code ZRATE, later ZWRATE and ROKANC, became the center 
of the Standard Model library of ZFITTER.
30 years later, at this year's Loops and Legs conference, I presented an electroweak Standard Model two-loop calculation for $Z \to 
{\bar b}b$, 
performed together with my coauthors. 
This project started in 2012 at a meeting at the Max-Planck Institute in Munich, where I met Ayres Freitas.
At that meeting the second-oldest speaker was about 20 years younger than me \ldots
~Our numerical fitting formula for the $Z{\bar b}b$ weak mixing angle will, presumably, get included into software packages like ZFITTER. 
During work on this report on the  $Z{\bar b}b$ project  I understood that there is a 
deep connection to one of my other hobbies -- the S-matrix approach to the $Z$ resonance.
This connection will be described shortly in the introduction, although it was not part of the oral presentation at the conference. 
Its understanding certainly will help to create a strict one per mille analysis tool for the $Z$ resonance as it is assumed to be needed 
for the next $e^+e^-$ collider, and perhaps at the LHC.  

I would like to thank 
my collegues for a decade-long, competitive, but also collaborative work in the research field of elementary 
particle physics, notably Arif Akhundov, Dima Bardin, Penka Christova, Dietmar Ebert, Jochem Fleischer, Ayres Freitas, Janusz Gluza, 
Wolfgang Hollik, Lida Kalinovskaya, Max Klein, Arnd Leike, Gottfried Mann, Sven Moch, Sabine Riemann, 
as well as 
Frank Kaschluhn, Karl Lanius and Paul S{\"o}ding for support.
With my PhD students Diet\-rich, 
Mark, Jochen, Alejandro, Valery (see 
\href{https://www.genealogy.math.ndsu.nodak.edu/id.php?id=29907}
{https://www.genealogy.math.ndsu.nodak.edu/id.php?\linebreak[4]id=29907}), and now Johann and Ievgen, work was and is  pleasure.   
In February 2015 I underwent a medical surgery, and I thank Professor Dr. med. habil. Ahmed Magheli from Charite in Berlin that I 
could afterwards continue working and successfully apply as a Fellow of the Polish Alexander von Humboldt Research Scholarship 2015, with 
host Janusz Gluza at the Silesian University at Katowice.

\medskip

Tord Riemann, 21 October 2016

\section{Introduction}
The study of the $Z$ boson resonance in $e^+e^-$ annihilation,
\begin{eqnarray}
\label{eq-1}
e^+e^- \to (\gamma,Z) \to f^+f^- ~(+ ~n\gamma)
,
\end{eqnarray}
has been performed with high precision at LEP and is planned with better precision at future $e^+e^-$ colliders.
Correspondingly, the theoretical predictions in the Standard Model are needed with 2-loop accuracy in the weak sector, and even better for 
QED and QCD.
Usually, the theoretical analysis is based not on cross sections as measured in reaction \eqref{eq-1}, including non-observed additional 
photons (and gluons),  but on so-called pseudo-observables, 
corresponding to 
\begin{eqnarray}\label{eq-2}
e^+e^- \to (\gamma,Z) \to f^+f^-
,
\end{eqnarray}
or even to the simpler reaction
\begin{eqnarray}\label{eq-221}
Z \to f^+f^- ~(+ ~n\gamma)
.
\end{eqnarray}
This is quite similar to the analysis of LHC events, which one tries to focus on the underlying hard $2\to 2$ process.
As a result, the analysis of observables rests on two relatively independent steps:
\begin{itemize}
 \item Unfolding the observed cross sections and representing them as $2\to 2$ (pseudo-) observables; the theoretical frame has to be 
sufficiently general in order not to bias step 2.
 \item Confronting the pseudo observables with specific theory predictions.
 \end{itemize}
 A key element of the theoretical analysis is the $Z{\bar b}b$ vertex, whose prediction at  two electroweak loops is the subject of our 
study.
 Two pseudo-observables are related to this vertex:
 \begin{itemize}
 \item The partial decay width $\Gamma(Z\to {\bar b} b)$.
 \item The decay asymmetry, related to the parameter $A_b$; we give its definition below in \eqref{eq-ab0}.
  \end{itemize}
The partial decay width $\Gamma(Z\to {\bar b} b)$  is related to the peak cross-section $\sigma_T$ of $e^+e^- \to (\gamma,Z) \to b^+b^- (+ 
n\gamma)$, also to the pseudo-cross section of
$e^+e^- \to (\gamma,Z) \to b^+b^-$, and $A_b$ is related to the angular asymmetry of the cross-sections.

In Born approximation everything looks relatively easy.
Neglecting here photon exchange, and  using a Breit-Wigner resonance $Z$ propagator with a priori mass $M_Z$ and  width $\Gamma_Z$ of the 
$Z$ boson, 
one derives a qualitatively good description of the $Z$ line shape close to the peak: 
\begin{eqnarray}
 \frac{d\sigma^B}{d\cos\theta} &\sim& G_F^2 ~
 \left|\frac{s}{s-M_Z^2+iM_Z\textcolor{black}{\Gamma_Z}}\right|^2
 \\\nonumber
&&\times ~ \left[(a_e^{B\;2}+v_e^{B\;2}) (a_b^{B\;2}+v_b^{B\;2})(1+\cos^2\theta) + (2a_e^Bv_e^B)(2a_b^Bv_b^B)(2\cos\theta) \right]
.
\end{eqnarray}
Symmetric or anti-symmetric integration over $\cos\theta$ allows to determine the two independent contributions.
One of them is the total cross section,
\begin{eqnarray}
 \sigma_T^B &\equiv& \int_{-1}^{1}d\cos\theta \frac{d\sigma^B}{d\cos\theta}
\sim
 \left|\frac{s}{s-M_Z^2+iM_Z\textcolor{black}{\Gamma_Z}}\right|^2 G_F^2\left(a_e^{B\;2}+v_e^{B\;2}\right) 
 \color{black}{ \left(a_b^{B\;2}+v_b^{B\;2}\right)}
\sim {\Gamma_{e}^B}  \Gamma_{b}^B
,
\end{eqnarray}
and the other one the forward-backward asymmetry, 
 \begin{eqnarray}
  \sigma_{FB}^B &\equiv& 
 \left[\int_{0}^{1}- \int_{-1}^{0}\right]
 d\cos\theta \frac{d\sigma^B}{d\cos\theta}
 \sim
 \left|\frac{s}{s-M_Z^2+iM_Z\textcolor{black}{\Gamma_Z}}\right|^2 G_F^2
( 2a_e^Bv_e^B) ~(2a_b^Bv_b^B)
,
\\
A_{FB}^B &\equiv& 
\frac{\sigma_{FB}^B} {\sigma_T^B} 
= 
\frac{3}{4} ~ \frac{2a_e^Bv_e^B}{a_e^{B\;2}+v_e^{B\;2}} 
~ \color{black}{ \frac{2a_b^Bv_b^B}{a_b^{B\;2}+v_b^{B\;2}}}  
 \equiv \frac{3}{4} ~ {A_e^B} ~ {A_b^B}
.
\end{eqnarray}
We observe the factorization of $\sigma_T^B$ into the product of two partial widths,  
\begin{eqnarray}
 \Gamma_{f}^B = \frac{G_F M_Z^3}{\sqrt{2}~6\pi} c_f \left(a_f^{B\;2}+v_f^{B\;2}\right),
\end{eqnarray}
and of $A_{FB}^B$ into the product of  two asymmetry functions,
\begin{eqnarray}
\label{eq-ab0}
 A_f^B = \frac{2a_f^Bv_f^B}{a_f^{B\;2}+v_f^{B\;2}}.
\end{eqnarray}
In Born approximation it is $a_f^B=\pm \frac{1}{2}$, $Q_e=-1$, $c_f$ the color factor, and 
\begin{eqnarray}
\frac{v_f^B}{a_f^B} = 1-4|Q_f|\sin^2\theta_W.
 \end{eqnarray}
The vector and axial vector couplings will get loop corrections, which may be calculated from the vertex diagrams ${\bf 
V}^{Zf\bar{f}}_\mu(k^2)$; for the 
$Z{\bar b}b$-vertex:
\begin{eqnarray}
\label{eq-ab}
g_{\rm V}^b(k^2)|_{k^2=M_Z^2} &=& \frac{1}{2(2-D)k^2} \, {\rm Tr}[\gamma^\mu \, \cancel{p}_1 \,
{\color{black}{\bf V}^{Zb\bar{b}}_\mu} \, \cancel{p}_2]|_{k^2=M_Z^2}, 
\\\label{eq-abaxi}
g_{\rm A}^b(k^2)_{k^2=M_Z^2} &=& \frac{1}{2(2-D)k^2} \, {\rm Tr}[\gamma_5 \,
\gamma^\mu \, \cancel{p}_1 \,
{\color{black}{\bf V}^{Zb\bar{b}}_\mu} \, \cancel{p}_2]_{k^2=M_Z^2}.
\end{eqnarray}
Here, we relate vertex corrections to effective couplings. In reality, realistic cross sections are measured, and one has to relate their 
couplings to  $g_{\rm V}^b(k^2)$ and $g_{\rm A}^b(k^2)$.

\medskip

Fitting programs like Gfitter are relating ``experimental'' values of e.g. $\Gamma_{b}$, $A_b$ with their theoretical predictions, e.g. in 
the 
standard model \cite{gfitter-baak:2015}.
But does this fit the original pseudo-observables?
To some approximation, it does, as can be seen in the Born formulae.
But one has to control the quantum corrections safely.
We know since long how to relate pseudo-observables in a strict way to the loop corrections 
\cite{Bardin:1988by,Bardin:1989tq,Bardin:1992jc,Bardin:1999yd,Akhundov:2014era}.
The amplitude for $e^+e^-$ annihilation into two massless fermions, and we assume here the final state to be massless, may be described 
to all orders of perturbation theory by four complex-valued form factors, which depend on the masses and the invariants $s$ and $t$, and  
which are chosen here to be $\rho_{ef}, \kappa_e, \kappa_f, \kappa_{ef}$; we 
quote  
from \cite{Bardin:1999yd}, eq. (3.3.1):\footnote{The left-projector is in that notations $L=(1+\gamma_5)/2$, while it is usually 
$L=(1-\gamma_5)/2$. We further stress that the notation covers all kinds of contributions, including also box diagrams.}
\begin{eqnarray}  
{\cal A}^{\sss{eff}}_{\sss{\zb}}(\sman,\tman)&\sim& 
        \ib\,e^2\,4\,\tcie\tcif\frac{\chi_Z(s)}{s}
        {\rho_{ef}(\sman,\tman)}
        \biggl\{
        \gadu{\mu}{\lpar 1+\gfd \rpar }
        \otimes \gadu{\mu} { \lpar 1+\gfd \rpar}    
              \\ \nonumber            &&
-4 |\qe | \stws {\kappa_e(\sman,\tman)}
        \gadu{\mu} \otimes \gadu{\mu}{\lpar1+\gfd\rpar}
       -4 |\qf | \stws {\kappa_f(\sman,\tman)}
        \gadu{\mu} {\lpar 1+\gfd \rpar } 
        \otimes \gadu{\mu}                              
             \\ \nonumber      &&
+
16 |\qe \qf| \stwf {\kappa_{ef}(\sman,\tman)}
        \gadu{\mu} \otimes \gadu{\mu} \biggr\}.
\label{processrhokappadef}
\end{eqnarray}
We use the definitions
\begin{eqnarray}
 \chi_Z(s) &=& \frac{G_FM_Z^2}{\sqrt{2}~2\pi\alpha}~\rho_Z(s),
 \\
\rho_Z(s) &=& \frac{s}{s-M_Z^2+iM_Z\textcolor{black}{\Gamma_Z(s)}}
 .
\end{eqnarray}
For the complete amplitude one sums over all relevant diagrams so that the form factors are perturbative series:
\begin{eqnarray}
 \rho_{ef} &=& 1 + \delta\rho_{ef} = 1+ \frac{\alpha}{\pi} \delta \rho_{ef}^{(1)} + \cdots,
 \\
 \kappa_{a} &=& 1 + \delta\kappa_{a} = 1+ \frac{\alpha}{\pi} \delta \kappa_{a}^{(1)} + \cdots, ~~~ a=e,f,ef
 .
\end{eqnarray}
Compared to a ``naive'' notation, we split from the rest of the amplitude the form factor $\rho_{ef}$ multiplicatively. If a diagram is 
represented by an original  set $\{\rho_{ef},{\bar \kappa_{e}},{\bar \kappa_{f}},{\bar \kappa_{ef}}\}$, this yields re-definitions for 
all the $\kappa_{a}$:  
\begin{eqnarray}
 \kappa_{a} = \frac{ {\bar \kappa_{a}}} {\rho_{ef}}, ~~~ a=e,f,ef
 .
\end{eqnarray}
The form factors, if introduced as it is done here, may be used for  definitions of an
effective Fermi constant and three effective weak mixing angles:
\begin{eqnarray}
G_{F}^{\rm eff} &=& \rho_{ef}(s,t) ~ G_{F},
\\
\sin^2 \theta_{W,e}^{\rm eff} &=& \kappa_e(s,t) ~ \sin^2 \theta_{W},
\\
\sin^2 \theta_{W,f}^{\rm eff} &=& \kappa_f(s,t) ~ \sin^2 \theta_{W},
\\
\sin^2 \theta_{W,ef}^{\rm eff} &=& \sqrt{\kappa_{ef}(s,t)} ~ \sin^2 \theta_{W},
\end{eqnarray}
where 
\begin{eqnarray}
\sin^2 \theta_{W} \equiv 1-\frac{M_W^2}{M_Z^2}.
\end{eqnarray}
The unique definition of an effective weak mixing angle is lost.

The Breit-Wigner propagator $\rho_Z(s)$ contains a width function which is predicted by perturbation theory. Its calculation deserves 
special attention, and for the moment the notation $\Gamma_Z(s)$  emphasizes that it originates from summing over self-energies like 
$\Sigma_Z(s)$.

The amplitude may be further rewritten, in order to introduce the familiar couplings $v_f, a_f$, which now will cover the loop 
corrections:
\begin{eqnarray}  
{\cal A}^{\sss{eff}}_{\sss{\zb}}(\sman,\tman)&\sim& 
        \ib\,e^2
        \frac{\chi_Z(s)}{s}
    a_e a_f       
        {\rho_{ef}(\sman,\tman)}
        \biggl(
%
              \gadu{\mu} \gfd \otimes \gadu{\mu} \gfd   
+ \frac{v_f}{a_f}        \gadu{\mu} \otimes \gadu{\mu} \gfd
+ \frac{v_e}{a_e}        \gadu{\mu} \gfd  \otimes \gadu{\mu}                              
+ \frac{v_{ef}}{a_ea_f}        \gadu{\mu} \otimes \gadu{\mu} 
        \biggr).
        \nonumber \\ 
\label{processrhokappadef-2}
\end{eqnarray}
Here, we made the choice that the axial couplings remain Born like, 
\begin{eqnarray}
 a_e &=& -\frac{1}{2},
 \\
 a_f &=& \pm\frac{1}{2}
 .
\end{eqnarray}
This choice means that the axial couplings remain to be real constants here, and that the (axial $\times$ axial) radiative corrections 
coming from a product of two vertices  like~\eqref{eq-abaxi} will be collected in the definition of $\rho_{ef}$, 
\begin{eqnarray}
 \delta \rho_{ef}^{ Z{\bar e}e Z{\bar b}b} &=& 
\frac{g_{\rm A}^e(k^2)_{k^2=M_Z^2}}{a_e} ~ \frac{g_{\rm A}^f(k^2)_{k^2=M_Z^2}}{a_f}
,
\end{eqnarray}
while the vector couplings are understood to contain radiative corrections.
From a final state $Z{\bar b}b$ vertex loop correction, one then gets e.g.:
\begin{eqnarray}
v_e &=&  v_e^B ,
\\
v_b &=& v_e^B + \delta  v_f^{Z{\bar b}b}     =  v_e^B + g_{\rm V}^f(k^2)_{k^2=M_Z^2},
 \\
v_{eb} &=& v_e^B \left(v_b^B +  \delta  v_{ef}^{Z{\bar b}b} \right)
       = v_e^B \left(v_b^B + \delta  v_f^{Z{\bar b}b}\right)
  .
\end{eqnarray}
From the above definitions we get  three relations between the vector couplings and the form factors~$\kappa$:
\begin{eqnarray}
 \frac{v_e}{a_e} &=& 1 -4 |\qe | \stws {\kappa_e},
 \\
 \frac{v_f}{a_f} &=& 1 -4 |\qf | \stws {\kappa_f},
 \\
 \frac{v_{ef}}{a_ea_f} &=& \frac{v_e v_f}{a_ea_f} + \Delta_{ef},
 \end{eqnarray}
 with
 \begin{eqnarray}
 \Delta_{ef} &=& 16 |\qe\qf | s_W^4 (\kappa_{ef}-\kappa_e\kappa_f)  
.
\end{eqnarray}
If $\kappa_{ef}-\kappa_e\kappa_f = 0$, there is factorization.
Factorization is broken by  photonic corrections and by box diagrams, while it is respected by weak vertex corrections and self-energies.

Having defined the amplitude, one may calculate, with standard text book methods,  a $2\to 2$ cross section. 
For unpolarized scattering one gets \cite{Bardin:1989tq}:
\begin{eqnarray}
\label{eq-sigma}
 \frac{d \sigma^{eff}}{d \cos\theta} = \frac{\pi \alpha^2}{2s} |\chi_Z(s)|^2
 \left[
(1+\cos^2 \theta) k_T + 2 \cos\theta k_{FB}
 \right]
 .
\end{eqnarray}
The symmetric part depends on 
\begin{eqnarray}
\label{eq-kt}
 k_T 
 &=& 
 |\rho_{ef}|^2 \left[ |a_ea_f|^2 + |v_ea_f|^2 + |a_ev_f|^2  + |v_{ef}|^2      \right]
 \\\nonumber
 &=& 
|\rho_{ef}|^2 |a_e|^2|a_f|^2\left[(1+|\frac{v_e}{a_e}|^2)(1+ |\frac{v_f}{a_f}|^2) 
+ \Delta_T 
\right]
 ,
 \end{eqnarray}
 with 
 \begin{eqnarray}
 \Delta_T &=& |\Delta_{ef}|^2 + 2 \Re e \left( \frac{v_e}{a_e} \frac{v_f}{a_f} \Delta_{ef}^*\right).
\end{eqnarray}
 Assuming factorization, this becomes 
 \begin{eqnarray}
 k_T 
 &=& 
  |\rho_{ef}|^2 \left[(|a_e|^2+|v_e|^2)(|a_f|^2+ |v_f|^2)\right],
  \end{eqnarray}
  and finally, neglecting additionally the  imaginary parts of $v_e$ and $v_f$ (and of $\Delta_T$): 
  \begin{eqnarray}
 k_T 
 &=& 
  |\rho_{ef}|^2 \left[(a_e^2+v_e^2)(a_f^2+ v_f^2)\right].
  \end{eqnarray}
  This is the formula usually applied to analyses.
  
  Similarly, for the anti-symmetric cross section part:
 \begin{eqnarray}
 \label{eq-kfb}
 k_{FB} 
 &=& 
 |\rho_{ef}|^2 a_e a_f \Re e  (v_e v_f^*+v_{ef})  
 \\\nonumber
 &=&
 a_ea_f
 \left(2
 \Re e \left[ \frac{v_e }{a_e} \right] \, 2\Re e \left[ \frac{v_f }{a_f} \right] + \Delta_{FB}\right)
,
 \end{eqnarray}
with
\begin{eqnarray}
 \Delta_{FB} &=& 2 \Re e \Delta_{ef} ,
\end{eqnarray}
 and after again neglecting non-factorizing terms and imaginary parts:
 \begin{eqnarray}
 k_{FB} 
 &=& 2 |\rho_{ef}|^2 
 (2a_ev_e)~(2a_fv_f) 
 .
  \end{eqnarray}
The cross section formula \eqref{eq-sigma} is the exact result from the amplitude square and averaging over the initial final helicity 
states. 

Here, a technical remark is at the place: 
Already in Born approximation, the photon exchange leads to non-factorization. 
It is numerically not small and has to be taken into account.
One may, formally, assume that the photon exchange Born amplitude is contained in the above-introduced four form factors. This was 
exemplified in \cite{Leike:1991if}.
Conventionally, one works with two interfering amplitudes as it is described in detail in the publications describing the ZFITTER project 
\cite{Bardin:1989di,Bardin:1989tq,Bardin:1992jc,Bardin:1999yd,Akhundov:2014era}.

Under the \textcolor{black}{assumption} that the form factors are independent of the scattering angle, we get for the total cross section 
and 
the forward-backward asymmetry: 
\begin{eqnarray}
  \sigma_{T} &=& \frac{4\pi \alpha^2}{3s} |\chi_Z|^2 ~ k_T ,
  \\
 \sigma_{FB} &=& \frac{\pi \alpha^2}{s} |\chi_Z|^2 ~ k_{FB} , 
  \end{eqnarray}
  and the forward-backward asymmetry becomes 
  \begin{eqnarray}
  A_{FB}^{eff} &=& \frac{3}{4} ~ \frac{k_{FB}}{k_{T}} .
\end{eqnarray}
If the form factors depend on the scattering angles, as it is the case for corrections from box diagrams, one has to study the 
numerical effect of that.  

Further observables may be introduced for polarized scattering, where the amplitude  \eqref{processrhokappadef} is taken 
between helicity 
projected states. 
This may be easily investigated following \cite{Bardin:1989tq}.

The loop-corrected asymmetry parameter $A_b$ as defined in \eqref{eq-ab0} will be set in relation to loop-corrected pseudo-observables at 
$s=\MZ^2$, in terms of the 
angular integrals $\sigma_{FB}, \sigma_T$ as defined in \eqref{eq-kt} and  \eqref{eq-kfb}:
\begin{eqnarray}
 \label{eq-afbpseudo2}
A_{FB}^{{\bar b}b} &=& \frac{\sigma_{FB}}{\sigma_T}
\\\nonumber&=& 
\frac{3}{4} ~
\frac{\Re e[2a_ev_e ~ 2a_bv_b + 4\sin^2\theta_W |Q_eQ_b|^2(\kappa_{eb}-\kappa_e\kappa_b) ]}
{ |a_ea_b|^2 + |v_ea_b|^2 + |a_ev_b|^2 +|v_{eb}|^2 }
 ~~  + \textrm{~~ corrections}
\\\nonumber
&=& \frac{3}{4} ~ A_\Pe ~ A_\Pb ~~ + \textrm{~~ corrections}.
\end{eqnarray}

The first ``corrections'' are due to neglected angular dependences of the form factors, and the second ``corrections'' are due to neglected 
non-factorizations and imaginary parts.

As discussed in detail in \cite{Awramik:2006uz}, 
as well as in earlier work \cite{%
Stuart:1991xk,
Veltman:1992tm
},
the weak mixing angle $\seff{b}$ and $A_\Pb$ are determined from the residue $R$ of the leading part 
of the resonance matrix element ${\cal \bar M}$ \eqref{eqsmatrix}. 
This residue may 
be determined in a very good and controlled approximation  from the 
renormalized vector and axial vector couplings of the vertices 
$V^{Ze\bar{e}}_\mu$ and $V^{Zb\bar{b}}_\mu$. 
To do so, we have to understand how the form factors are composed.
Besides the  terms from $s$ channel $Z$ boson exchange, they contain terms from $s$ channel photon exchange, from box diagrams (with 
weak bosons, but also with photon exchanges), vertices, 
self-energies.\footnote{We remark here that not only self-energies and vertices, but also arbitrary box diagrams may be inserted exactly 
into the form factors \cite{Bardin:1988by,Bardin:1989tq,Bardin:1999yd}.}
Some of these terms are enhanced by the resonance form of the transition, others are not.
One has to understand some summation of terms which otherwise would explode at $s=M_Z^2$.

We will now discuss shortly the consequences of the fact 
that we are studying a resonance.
Here arises the question, what is the correct and model-independent formulation of the amplitude from the 
point of view of 
general quantum field theory?
In order to respect general principles -- unitarity, analyticity, gauge invariance -- one may use the pole scheme \cite{Stuart:1991xk}.
In the pole scheme, one makes the following ansatz for the 
amplitude ${\cal \bar M}$ as a function of the scattering energy, or as a function of the corresponding relativistic invariant $s$. In a 
sufficiently small neighborhood around the pole position, it is a Laurent expansion  with position of the pole $s_0$ defined by the mass 
$m_Z$ and its width $g_Z$, 
\begin{eqnarray}
 s_0 &=& m_Z^2 -i m_Z g_Z
 ,
\end{eqnarray}
and the residue $R$, plus a background term $B$. The latter is a Taylor expansion:
\begin{eqnarray}
\label{eq-mgeneric}
\label{eqsmatrix}
 {\cal \bar  M} \sim \frac{R}{s-m_Z^2+im_Z g_Z} + \sum_{n=0}^{\infty} \frac{b_n}{s_0} \left(1-\frac{s}{s_0}\right)^n 
 .
\end{eqnarray}
A kind of master formula for $R$ is equation (12), together with (13), of \cite{Awramik:2006uz,Veltman:1992tm}:
\begin{align}
R &= 
 \begin{aligned}[t]
 &z_\Pe^{(0)} \, R_{ZZ} \, z_\Pf^{(0)} + \left[ \hat{z}_\Pe^{(1)}(\MZ^2)\, z_\Pf^{(0)} +
   z_\Pe^{(0)} \, \hat{z}_\Pf^{(1)}(\MZ^2) \right] 
   \left[ 1+ {\Sigma_{Z Z}^{(1)}}'(\MZ^2) \right] 
   \\
 &+ \hat{z}_\Pe^{(2)}(\MZ^2) \, z_\Pf^{(0)} +
   z_\Pe^{(0)} \, \hat{z}_\Pf^{(2)}(\MZ^2)
   + \hat{z}_\Pe^{(1)}(\MZ^2) \, \hat{z}_\Pf^{(1)}(\MZ^2) \\
 &- i \MZ \GZ \left[ \mbox{$\hat{z}_\Pe^{(1)}$}'(\MZ^2)\, z_\Pf^{(0)} +
   z_\Pe^{(0)} \, \mbox{$\hat{z}_\Pf^{(1)}$}'(\MZ^2) \right],
 \end{aligned} \\
R_{ZZ} &= 
 \begin{aligned}[t]
 &1 - {\Sigma_{Z Z}^{(1)}}'(\MZ^2) \\
 &- {\Sigma_{Z Z}^{(2)}}'(\MZ^2) + \left( {\Sigma_{Z Z}^{(1)}}'(\MZ^2) \right)^2
   + i \MZ \GZ \, {\Sigma_{Z Z}^{(1)}}''(\MZ^2) \\
 &- \frac{1}{\MZ^4} \left( {\Sigma_{\gamma Z}^{(1)}}(\MZ^2) \right)^2
   + \frac{2}{\MZ^2} \, {\Sigma_{\gamma Z}^{(1)}}(\MZ^2) \,
     {\Sigma_{\gamma Z}^{(1)}}'(\MZ^2).
 \end{aligned}
\end{align}
The $\Sigma_{V_1V_2}, V_i=Z,\gamma$ stand for self-energies, and the vertices are defined there as
\begin{eqnarray}
\Gamma[Z_\mu f\bar{f}] &\equiv&
 z_{\Pf,\mu} = i \gamma_\mu (v_\Pf+a_\Pf \gamma_5), 
 \\
 \Gamma[\gamma_\mu f\bar{f}] &\equiv&
 g_{\Pf,\mu} = i \gamma_\mu (q_\Pf+p_\Pf \gamma_5), 
 \\
\hat{z}_{\Pf,\mu}(k^2) &=& i \gamma_\mu 
\left[\hat{v}_\Pf(k^2)+\hat{a}_\Pf(k^2) \gamma_5\right] 
\nonumber\\
&\equiv& i \gamma_\mu \left[v_\Pf(k^2)+a_\Pf(k^2) \gamma_5\right] 
     - i \gamma_\mu \left[q_\Pf(k^2)+p_\Pf(k^2) \gamma_5\right]
\frac{\Sigma_{\gamma Z}(k^2)}{k^2 + \Sigma_{\gamma \gamma}(k^2)} 
.
\end{eqnarray}
For details of notations, we refer to \cite{Awramik:2006uz}.

Here, ${\cal \bar  M}$ stands for the functional form of the  amplitude introduced in \eqref{eq-ab}. 
Within our formalism, one may write in full generality:
\begin{eqnarray}
 \rho_{ef} &=&
 \frac{R_r}{s-s_0} + \sum_{n=0}^{\infty} \frac{b_{r,n}}{s_0} \left(1-\frac{s}{s_0}\right)^n 
 ,
 \\
 \kappa_e &=&
\sum_{n=0}^{\infty} \frac{b_{e,n}}{s_0} \left(1-\frac{s}{s_0}\right)^n 
,
 \\
  \kappa_f &=&
\sum_{n=0}^{\infty} \frac{b_{f,n}}{s_0} \left(1-\frac{s}{s_0}\right)^n 
,
 \\
 \kappa_{ef} &=&
\sum_{n=0}^{\infty} \frac{b_{ef,n}}{s_0} \left(1-\frac{s}{s_0}\right)^n 
.
\end{eqnarray}
Because $\rho$ is chosen to be an overall factor, it is appropriate to include the resonating part of the amplitude here.\footnote{
One might, instead, hold a resonating overall factor of the amplitude $\chi_Z(s)$ outside the form factors. This is done in ZFITTER. Then 
$\rho$ has to be understood as a Taylor series, and if a specific contribution is non-resonating, e.g. because it is due to photon 
exchange, the first coefficient of $\rho$ would vanish, $b_{r,0}=0$ (see \cite{Leike:1991if} for more details).}
From the generic formula \eqref{eq-mgeneric}, one gets all the expressions for $\sigma_T, A_{FB}, A_{LR}$ etc. as explained above 
\cite{Leike:1991pq,Riemann:1992gv}. As a result, all these 
quantities $\sigma_A$ have the same form, but depend on different terms $R_{A,r}$ and $b_{A,f,n}$, which are bi-linear compositions of 
the coefficients $R_{r}$ and $b_{f,n}$.

The Breit-Wigner function used here deviates from the Breit-Wigner function as it was used by the  LEP collaborations, where 
$\Gamma_Z(s)= s/M_Z^2 ~ \Gamma_Z$ was used instead of $\Gamma_Z(s)=g_Z$.
The difference is not negligible and amounts to \cite{Bardin:1988xt}:
\begin{eqnarray}
 m_Z &=& \frac{M_Z}{\sqrt{1+\Gamma_Z^2/M_Z^2}} ~\approx~  M_Z - \frac{1}{2}\Gamma_Z^2/M_Z  ~\approx~ M_Z - 34 ~\mathrm{MeV}
,
 \\
 g_Z &=& \frac{\Gamma_Z}{\sqrt{1+\Gamma_Z^2/M_Z^2}} ~\approx~  \Gamma_Z - \frac{1}{2}\Gamma_Z^3/M_Z^2  
 ~\approx~ \Gamma_Z - 1 ~\mathrm{MeV} 
.
 \end{eqnarray}
We have expansions both around the pole position $s_0$ and in the coupling constants $\alpha$ and $\alpha_s$, and have to assume that 
$\alpha, \alpha_s$ and $g_Z/m_Z$ and also $1-s/m_Z^2$ are of the same numerical order.
As a consequence, in an electroweak calculation, $R$ is needed to order ${\cal O}(\alpha^2)$, the coefficients $b_0$ to order ${\cal 
O}(\alpha)$, and the $b_1$ etc. to leading order only.
One has to observe that also the quantity $g_Z$ itself is a prediction of the theory, beginning at  order ${\cal O}(\alpha^2)$.

A further complication comes from the fact that there are not so small higher order photonic corrections.
There are two approaches to that.
Either one assumes the photon exchange amplitude as a separate quantity, which interferes with the $Z$ boson amplitude, and takes this 
correctly into account.
This was done in the ZFITTER approach \cite{Bardin:1989tq,Bardin:1999yda,Akhundov:2014era}.
To the perturbative orders covered by ZFITTER, this was a controlled approach.
In general, it might be more consistent to work with only one amplitude and to understand photonic corrections a a part of background. 
Then, nevertheless it makes sense to calculate those parts of the photonic backgound with a precision needed by experiment, and to separate 
this from the unknown parts of the background, as was discussed in \cite{Riemann:2015wpn}. 
The above considerations help to understand the hierarchy of corrections.
Weak vertex corrections as well as weak self-energies contribute to $R$, while all the photonic corrections and also the box diagrams go 
into the background $B$.
This means that a two-loop calculation for the $Z$ resonance has to include only vertices and self-energies at two loops -- 
these are the factorizing corrections.
An immediate consequence is that for the calculation of an asymmetry like $A_{FB}$ close to the $Z$ peak, 
one needs only the $v_e/a_e$ and $v_f/a_f$, derived from (self-energy- and) vertex corrections, at two loops, and the other terms with less 
accuracy.\footnote{Strictly speaking, the $A_f$ is dependent on the scattering channel 
for which it is measured or calculated. Using e.g. $A_e$ as it is measured from muon pair production for the determination of $A_b$ from  
$A_{FB}$ as it is measured from ${\bar b}b$ production, one has check that this is consistent to the accuracy aimed at. }
The photonic corrections, as well as the box terms are not resonating and thus suppressed compared to the resonance residue. 
As a consequence, all the complicated two-loop boxes are negligible here, while the photonic corrections and the one-loop box terms are 
well-known and may be considered as a correction. In ZFITTER, this is organized in the various interfaces 
\cite{Bardin:1992jc,Bardin:1999yda}.
The numerical details have been carefully studied in \cite{Bardin:1999gt}, and never again since then.

We come now back to the definition of the effective weak mixing angle:
\begin{eqnarray}
 \seff{f} 
 &\equiv& \sin^2\theta_W ~ \Re e ~\kappa_f =  \frac{1}{4|Q_f|} \left(1-\Re e ~\frac{v_f}{a_f}\right)
 .
\end{eqnarray}
This means also 
\begin{eqnarray}
  \seff{f}
  &=& \left(  1-\frac{M_W^2}{M_Z^2}  \right) \left( 1+ \Delta \kappa_f \right)
  .
\end{eqnarray}
According to its definition, the $A_f$ is a function of one variable only; for $b$-quarks:
\begin{eqnarray}
A_\Pb &\equiv& \frac{2\re \frac{g_{\rm V}^b}{g_{\rm A}^b}}{1+(\re \frac{g_{\rm V}^b}{g_{\rm A}^b})^2}
\\\nonumber
 &=& \frac{1-4|Q_b|\seff{b}}{1-4|Q_b|\seff{b}+8Q_b^2
 \bigl ( \seff{b} \bigr )^2}\,
\end{eqnarray}
The so far best measurement is due to LEP~1 measurements \cite{ALEPH:2005ablast}:
\begin{eqnarray}
A_\Pb &=&
 0.899 \pm 0.013 
 .
\end{eqnarray}
For the weak mixing angle, this means:
\begin{eqnarray}
 \seff{b}
 &=&
0.281 \pm 0.016 
.
\end{eqnarray}
This value corresponds to an experimental accuracy of about 5.7\%.
At the next lepton collider, one aims at electroweak per mille measurements, which motivates complete weak two-loop predictions. We will 
see later that this aim is achieved for $A_b$ with our new result. 

To summarize this part of the discussion:
The $Z{\bar b}b$ asymmetry parameter $A_b$ may be expressed by the effective weak mixing angle $\seff{b}$, or seen in a 
different way:
One may determine the effective weak mixing angle $\seff{b}$ from the asymmetry parameter $A_b$, which by itself may be determined 
experimentally from combinations of pseudo-observables, and theoretically from the ratio $v_b/a_b$.
Here, the vertex form factor $\rho_b$ drops out.
The calculation of all the relevant radiative loop corrections at two loop order or more is involved.
The relation of these radiative corrections to the $Zff$ width and asymmetry parameters is simple, while the relations of the various width 
and asymmetry parameters to realistic observables or to pseudo observables need a careful control.

What we did not discuss so far is the relation of ``true'', or ``realistic'', observables and pseudo-observables.
It is constituted by the determination of the $2\to 2$ hard scattering observable from the experimentally accessible 
cross sections with multi-particle final states, where the $2\to 2$ amplitudes contribute together with more complicated final states which 
may not be distinguished.
One has to cover additional soft photons, gluons, but also $e^+e^-$-pairs etc.

It is not the aim here to discuss this in detail.
Up to additional corrections, at the $Z$ resonance the bulk of realistic observables may be described theoretically as a folding of the 
pseudo observables with some kernel functions.
The experimentally accessible total cross section e.g. may be written as follows:
\begin{eqnarray}
\label{eq-t}
 \sigma_{T}^{exp} (s) &=& \int d\frac{s'}{s} ~ \rho_T\left(\frac{s'}{s}\right) \sigma_{T}(s') + \cdots 
 .
\end{eqnarray}
Similarly, and with the same kernel function, one may describe polarization and helicity asymmetries.
The notable exception is, due a different angular dependence, the forward-backward asymmetric cross section:
\begin{eqnarray}
\label{eq-fb}
\sigma_{FB}^{exp} (s) &=& \int d\frac{s'}{s} ~ \rho_{FB}\left(\frac{s'}{s}\right) \sigma_{FB}(s') + \cdots
 ,
\end{eqnarray}
where $\rho_{FB}(s'/s) \neq \rho_T(s'/s)$.
For the explicit expressions for $\rho_{FB}(s'/s)$ and $\rho_T(s'/s)$, as well for more involved contributions, see e.g. 
\cite{Bardin:1988ze,Bardin:1989cw,Bilenky:1989zg,Bardin:1990fu,Christova:1999cc,Jack:2000as}.
Because at the $Z$ resonance the soft photon radiation dominates over hard photon emission, and the radiator kernels 
$\rho_{FB}(s'/s)$ and $\rho_T(s'/s)$ start to deviate from each other for hard emissions, one may 
use $\rho_T(s'/s)$ in some approximation for the prediction of all the realistic observables:
At the resonance, hard radiative emissions are kinematically suppressed.  
But when non-resonating parts become important, one has to take notice of their difference.

As an instructive example, we reproduce here the ${\cal O}(\alpha)$ approximated radiator functions for initial state radiation
~\cite{Bardin:1989cw}:
\begin{eqnarray}
\label{eq-rhoT}
 \rho_T\left(\frac{s'}{s}\right) &\sim& H_e(v) 
= Q_e^2 \frac{\alpha}{\pi} \left( L_e-1 \right)\frac{1+(1-v)^2}{v},
\end{eqnarray}
compared to
\begin{eqnarray}
\label{eq-rhoFB}
 \rho_{FB}\left(\frac{s'}{s}\right) &\sim& h_e(v) 
= Q_e^2 \frac{\alpha}{\pi} \left( L_e-1 
 -\ln \frac{1-v}{ \left(1-\frac{v}{2}\right)^2}
 \right)
 \frac{1+(1-v)^2}{v} \frac{1-v}{\left(1-\frac{v}{2}\right)^2}.
\end{eqnarray}
Here it is $L_e=\ln (s/m_e^2)$ and $v=1-s'/s$. The $v$ vanishes in the soft photon limit,  and $\rho_{FB}$ approaches $\rho_T$.

The unfolding of realistic observables according to \eqref{eq-t} and \eqref{eq-fb} can be performed with the analysis tools TOPAZ0 
\cite{web-topaz0-torino:2016,Montagna:1993ai,Montagna:1995ja,Montagna:1998kp} and  ZFITTER. The latter one relies on the work quoted 
above for $\rho_{T}$ and $\rho_{FB}$.
Evidently, the result of unfolding depends on the model chosen for the hard process $\sigma^0_{T}(s')$ or $\sigma^0_{FB}(s')$.
This fact is reflected by the various model-dependent so-called interfaces of ZFITTER.

\medskip

Finally, the following question has to be answered:
How may one take into account the resonating Breit-Wigner form of the pseudo-observables when unfolding?
The answer is given by the so-called S-matrix approach to the $Z$ resonance.
This task was not covered in the original versions of ZFITTER. A relatively simple version was offered with the interface ZUSMAT of 
ZFITTER. This interface  was finally replaced by a call to the independent Fortran package SMATASY 
\cite{Leike:1991pq,Riemann:1992gv,Adriani:1993sx,SRiemannPhD:1994,Kirsch:1994cf,gruenewald-smatasy:2005,Riemann:2015wpn}.
It is not the intention here to describe details of the approach.
We only mention that one may introduce in \eqref{eq-t} and \eqref{eq-fb} the effective Born approximations as they are derived from 
amplitudes of the form \eqref{eqsmatrix}, where the form factors  are expressing the corresponding resonance parameters $R_A$, $s_{A,0}$ 
and $b_{A,n}$.
Consequently, unfolding allows the determination of $s_0$ and of certain combinations of $R_A$ and, depending on the experimental 
accuracy, also of the background parameters $b_n$.

The theoretical predictions in the approach are based on the radiator functions, but get modified.
For the realistic (unfolded) forward-backward asymmetry one derives equation (28) of \cite{Riemann:1992gv}:
\newcommand{\rgt}{r_{\gamma}^T}
\newcommand{\rt}{R_T}
\newcommand{\rfb}{R_{FB}}
\newcommand{\rpol}{R_{pol}}
\newcommand{\ra}{R_A}
\newcommand{\rza}{r_0^A}
\newcommand{\rzt}{r_0^T}
\newcommand{\ia}{J_A}
\newcommand{\roa}{r_1^A}
\newcommand{\rot}{r_1^T}
\newcommand{\iT}{J_T}
\begin{eqnarray}
{\bar A}^{FB}(s) = {\bar A}_0^{FB} + {\bar A}_1^{FB} 
\left(1-\frac{s}{s_0}\right)
+  {\bar A}_2^{FB}
             \left(1-\frac{s}{s_0}\right)^2 + \cdots
\label{e24}
\end{eqnarray}
with e.g.
\begin{eqnarray}
{\bar A}_0^{FB} = \frac{C_R^{FB}(s)}{C_R^{T}(s)}
   \frac{\rfb}{\rt + \cdots }
\approx
0.998 \frac{\rfb}{\rt + 0.001}
\label{e25}
\end{eqnarray}
and 
\begin{eqnarray}
C_R^{A}(s) = \int dk  \, \rho_A (k)
\left[ \frac{s'}{s} \, \,
\frac{|s-s_0|^2}
     {|s'-s_0|^2}      \right],
                              ~~~A=T,FB.
\label{e21}
\end{eqnarray}
Here, it is $k=\frac{s'}{s}$ and examples for the radiators $\rho_A$ are introduced in \eqref{eq-rhoT} and \eqref{eq-rhoFB}.
The unfolding has to be performed with the same precision as the interpretation of the pseudo observable, e.g. with account of two 
or more loops at the per mille level.

For the $b$-pair production, one has to derive from data e.g. the residues $R_T, R_{FB}$.
Their ratio $R_{FB}/R_T$  gives then (after unfolding and with the approximations mentioned)
\begin{eqnarray}
 \frac{R_{FB}}{R_T} &=& \frac{3}{4} A_e A_b,
\end{eqnarray}
and for known $A_e$ from other measurements one may derive $A_b$.

If one intends to create a modernized per-mille version of ZFITTER for a study of the $Z$ resonance at a future lepton collider, one has to 
foresee interfaces which carefully take into account the notations and concepts described here.  

Now we come back to the very determination of the weak bosonic two-loop corrections to $A_b$ and to $\seff{b}$.
%
Explicit generic formulae for the residue $R$ of the $Z$ resonance amplitude, respecting general principles -- unitarity, analyticity, gauge 
invariance --, 
are given with equations ~(12) and ~(13) of 
\cite{Awramik:2006uz}, with a reference to earlier work \cite{Veltman:1992tm}.

\section{\bf The $Z$-boson width}
\begin{figure}[t]
\begin{center}
\includegraphics[width=0.2\linewidth]{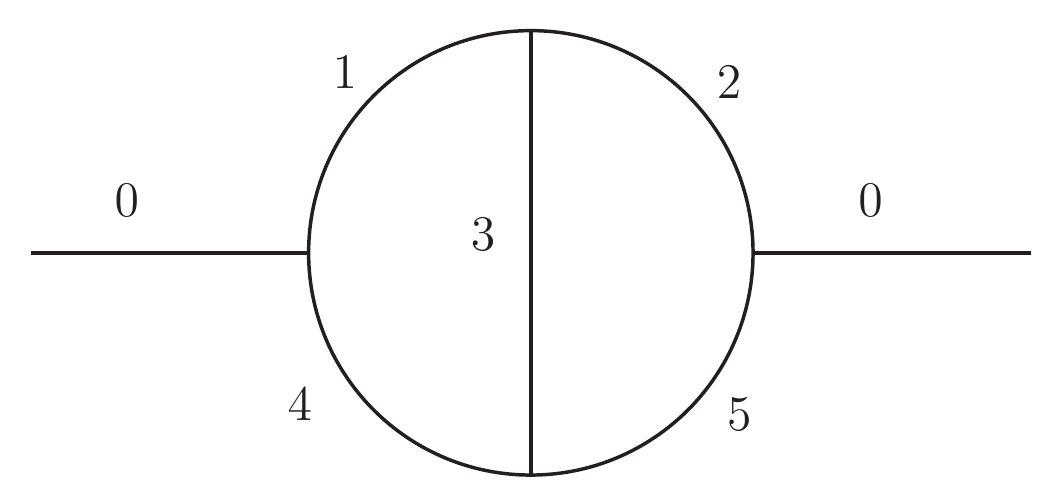}
\includegraphics[width=0.2\linewidth]{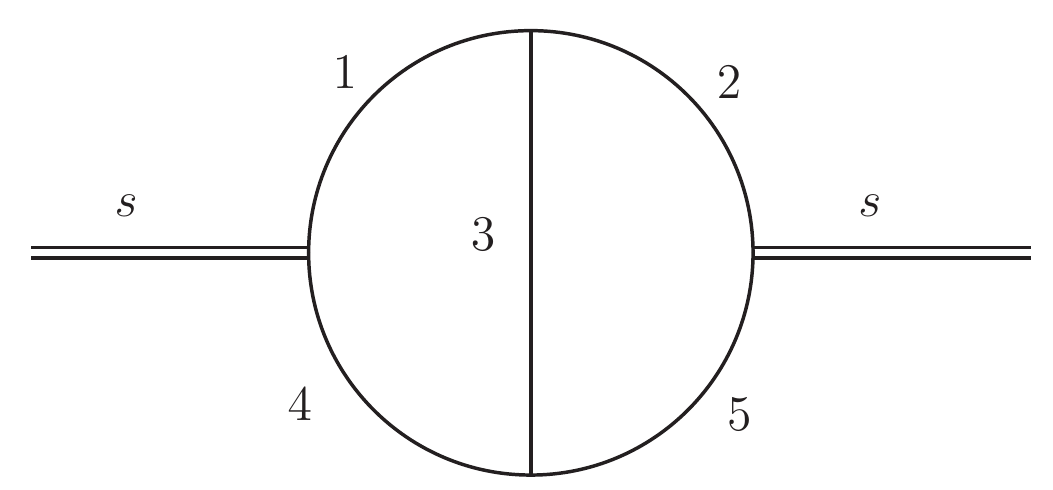}
\includegraphics[width=0.2\linewidth]{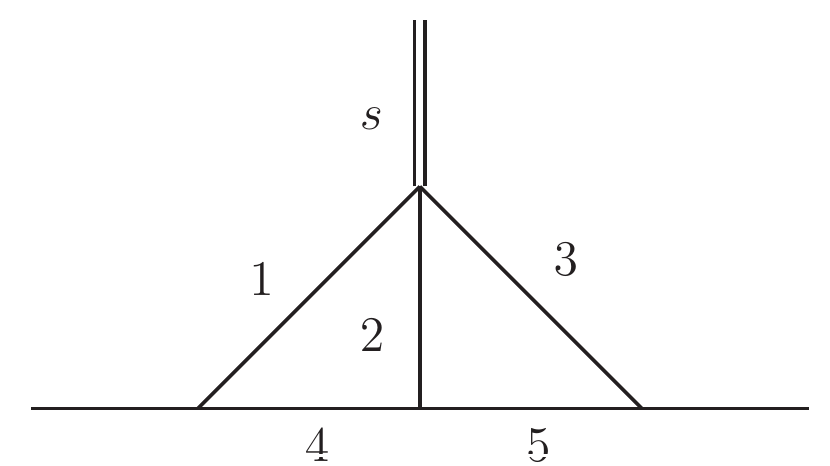}
\includegraphics[width=0.2\linewidth]{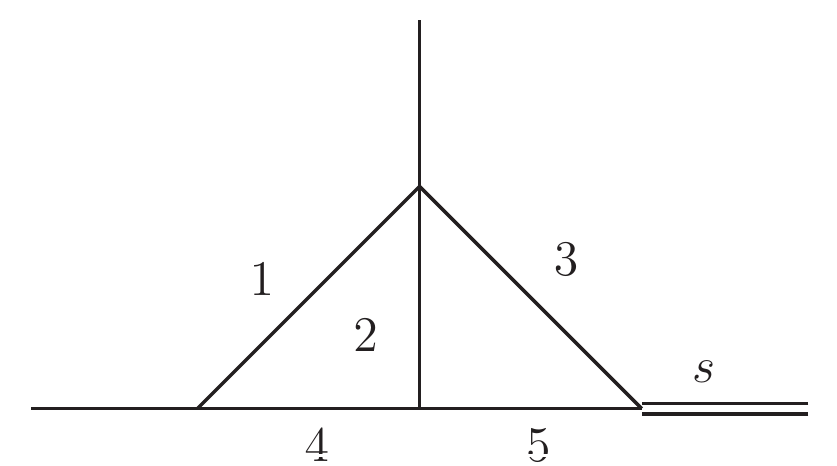}
\\
\includegraphics[width=0.3\linewidth]{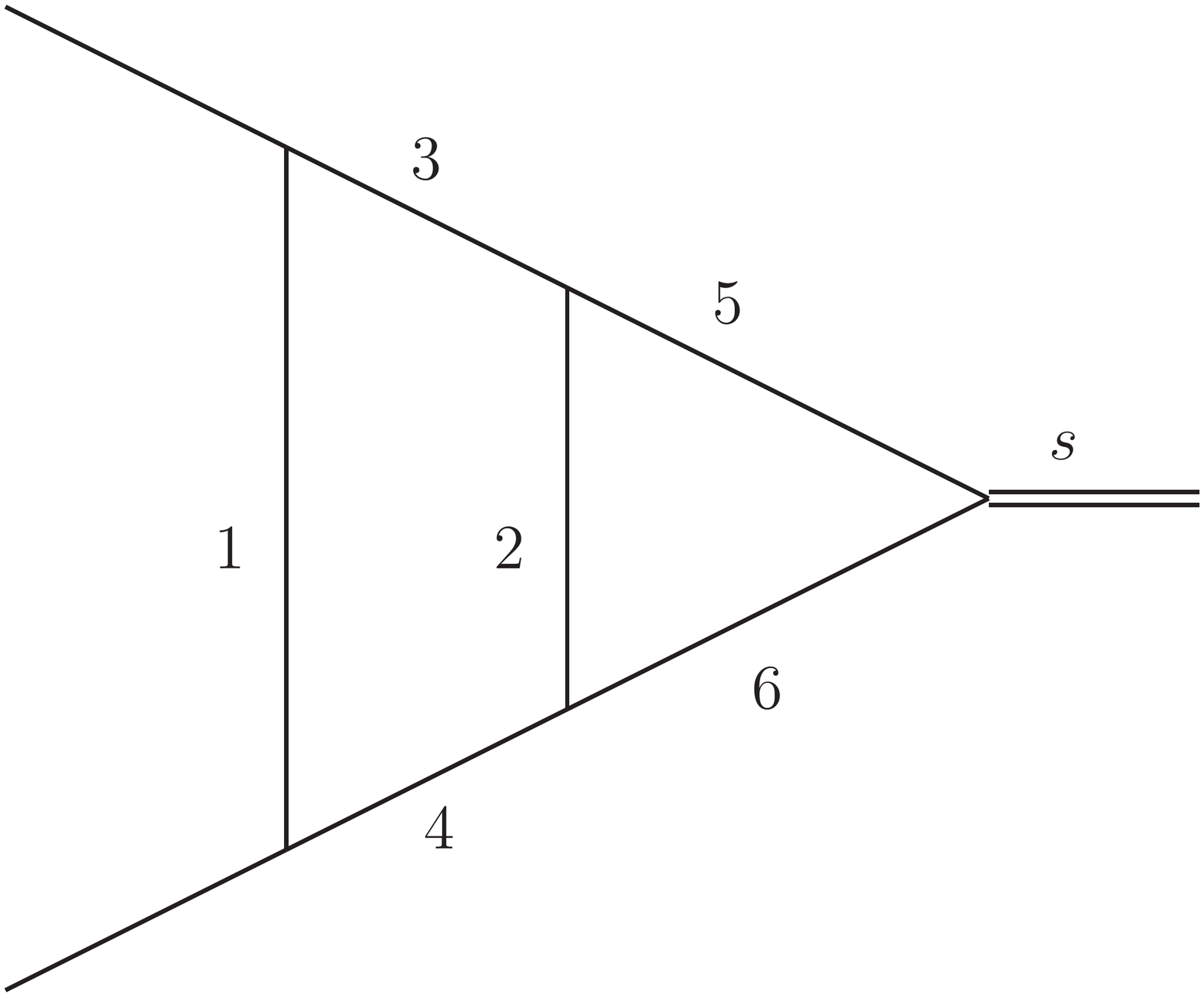} 
\includegraphics[width=0.3\linewidth]{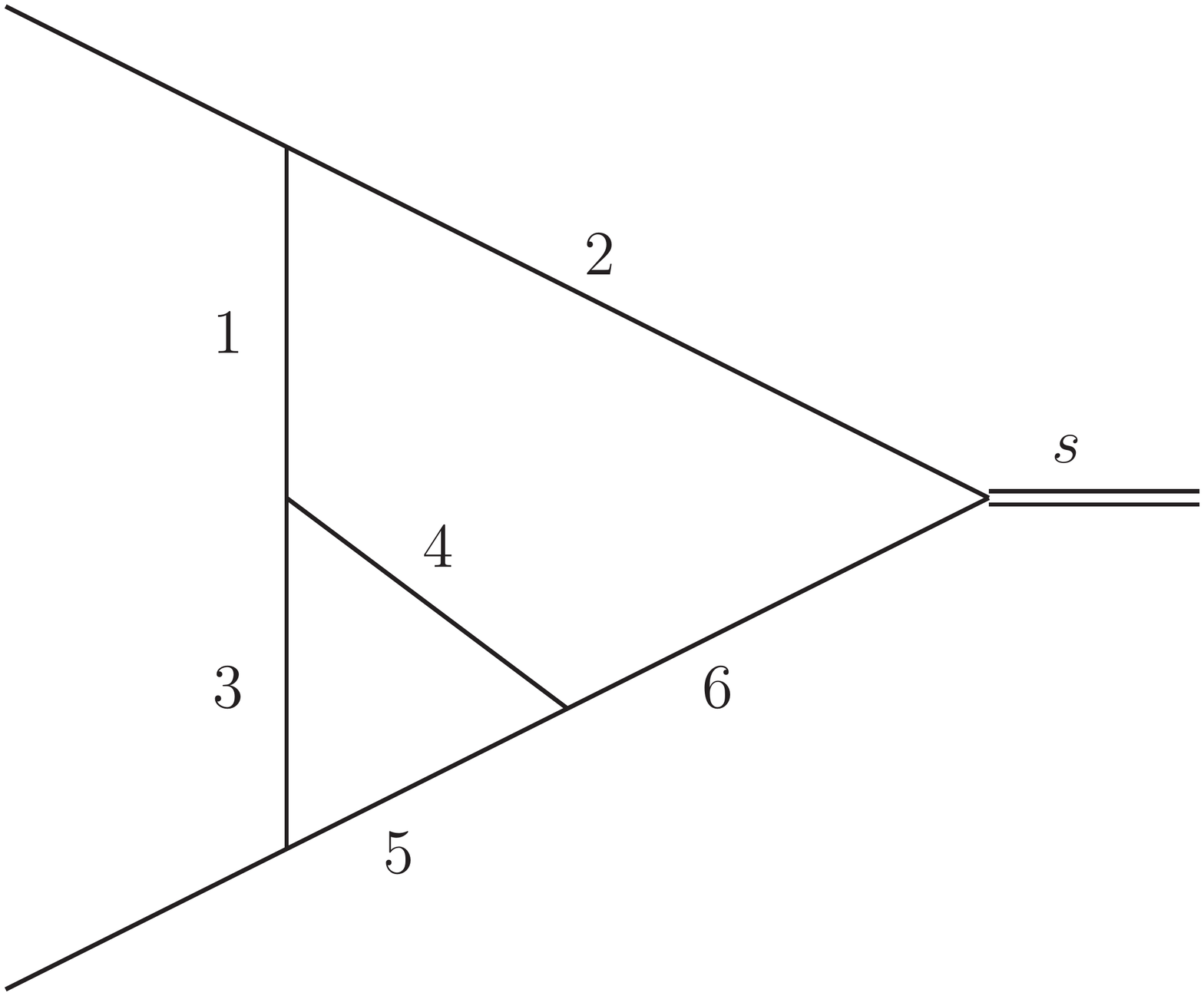} 
\includegraphics[width=0.3\linewidth]{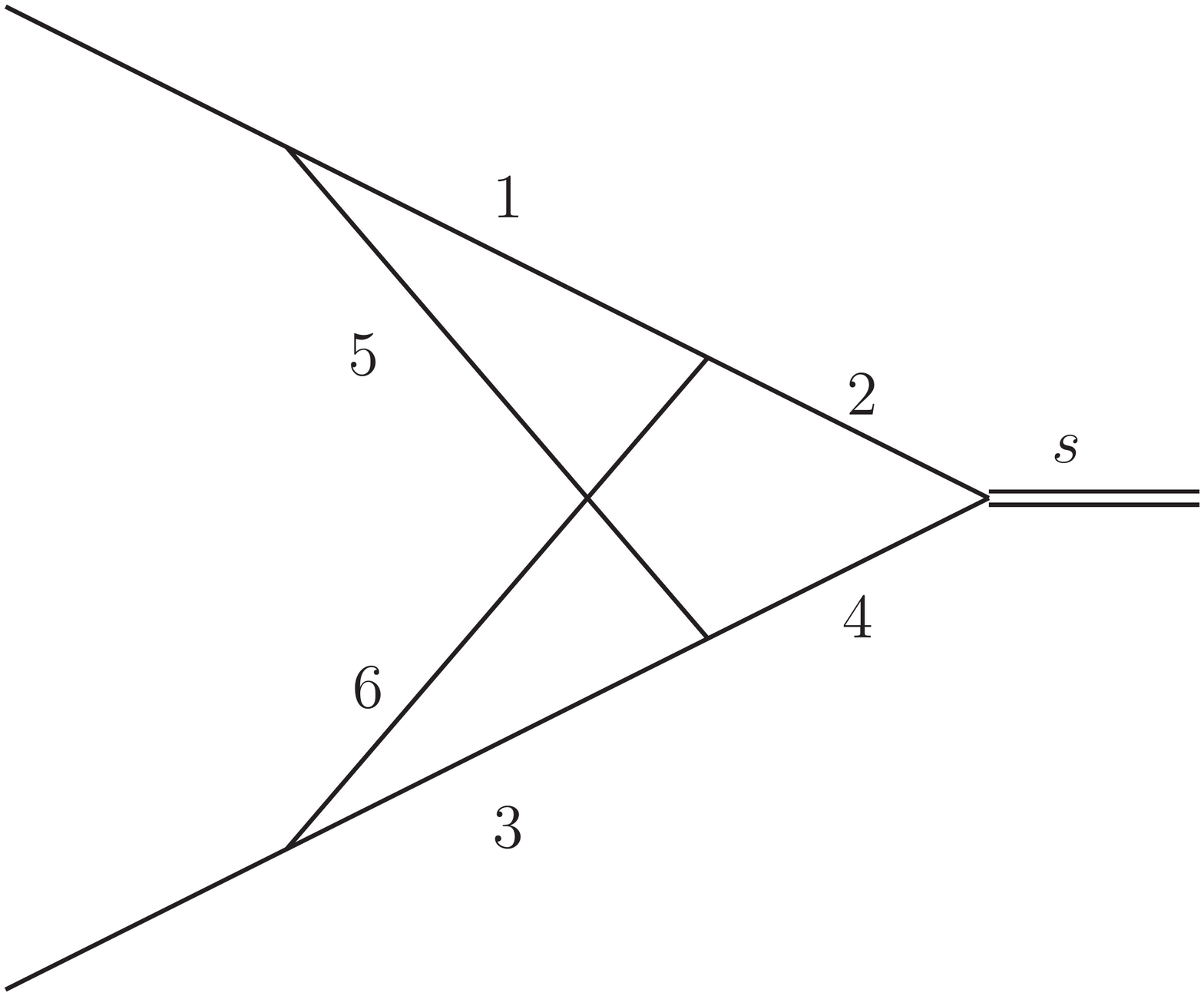} 
\end{center}
\caption{\label{fig-topo2}The bosonic electroweak two-loop topologies of the $Z{\bar b}b$ vertex.}
\end{figure}

As explained above, the $Z{\bar e}e$ and  $Z{\bar b}b$ vertices constitute the main contributions to the pseudo observables of the $Z$ 
resonance,
\begin{eqnarray}
 \Gamma(z\to {\bar b} b) &\sim& | {\cal M}_{1-loop} + {\cal M}_{2-loop} + \cdots|^2 + \cdots,
\\
{\cal M}_{2-loop} &\sim& \cdots + {\bar u}~ {\color{black}{\bf V}^{Z b{\bar b}}_{\mu}} u~ \epsilon^\mu + \cdots
\end{eqnarray}
The two-loop electroweak {\em fermionic corrections} to $A_b$ and $\sin^2 \theta^{\rm b \bar b}_{\rm eff}$ were determined in 
\cite{Awramik:2008gi}.
We have calculated now the so far unknown {\em bosonic integrals} for the 2-loop diagrams for the vertex 
{\color{black}{${\bf V}^{Zb{\bar b}}_\mu$}}. 
They include the topologies shown in figure \ref{fig-topo2}. 

\subsection{The electroweak one-loop corrections to the $Z$-boson vertex}  
Around 1980 it became evident that the Glashow/Salam/Weinberg model might become the electroweak Standard Model.
Consequently, some more 
elaborated loop calculations became meaningful.

Seen from today, the one-loop calculations of that time look quite simple.
One had to understand complex logarithms, the Euler dilogarithm $\mathrm{Li}_2$, and to read basically two seminal papers by  't Hooft and 
Veltman \cite{'tHooft:1978xw} and by Passarino and Veltman \cite{Passarino:1978jh}.
Among the first substantial, independent  electroweak projects was a study in the unitary gauge by the Dubna group, founded by Dima Bardin.
They studied complete electroweak radiative corrections for decays and scattering processes, including $e^+e^- \to {\bar f}f$ and $Z 
\to {\bar f}f$, assuming all fermions being massless.
In \cite{Bardin:1981sv}, no expanded numerics was performed.
Triggered partly by the detection of $Z\to l^+ l^-$ by the  UA1 and UA2 experiments at CERN, there were several calculations of $Z$-decay 
into leptons \cite{Consoli:1983yn,Fleischer:1984vt}, also under the assumption of massless fermions.
The decay $Z \to {\bar b}b$ gets contributions from $m_t$-dependent vertex contributions, and it is not covered by these calculations, 
if the top quark is heavy. The top quark mass was unknown at that time, but the 
experimental mass limits were growing up.
People at Dubna (Akhundov, Bardin, Riemann) observed that one can cover the amplitude with account of the additional $m_t$-dependent terms 
by adding up two known pieces:
\\
${\cal M}(Z\to {\bar q}q)$ with $m_q=0$,
and
\\
${\cal M}(Z\to {\bar q_1}q_2)$ with $q_1, q_2$ having the same isospin, but $q_1\neq q_2$.
\\
The first piece was known from \cite{Bardin:1981sv}, and the second one, non-vanishing only if a loop fermion mass is non-vanishing, from a 
study of flavor-nondiagonal $Z$ decays \cite{Mann:1983dv}.
So the two were combined and accomplished by an independent recalculation of the whole $Z{\bar b}b$ amplitude. The preprint  
JINR-E2-85-617 appeared in August 1985, and soon later the publication \cite{Akhundov:1985fc}.
The language of $\rho_Z$ and $\kappa_Z$ for the radiative corrections was used there, and the Fortran program {ZRATE} became the 
first piece of the electroweak Standard Model library of the ZFITTER project \cite{Akhundov:2014era,Bardin:1999yda,Arbuzov:2005ma}.
To give an example of the notation of form factors, we reproduce from \cite{Bardin:1992jc} the leading terms of the additional top quark 
corrections to the $Z{\bar b}b$ vertex compared to the $Z{\bar d}d$ vertex:
\begin{eqnarray}
\rho_b^Z
 = \rho_d^Z
  - 2 \frac{\Delta_b(m_t^2)}{a_b},
\label{delrhob}
\\
\kappa_b^Z
 = \kappa_d^Z
  +
  \frac{\Delta_b(m_t^2)}{a_b}.
\label{delkab}
\end{eqnarray}
The exact form factors have been implemented in ZFITTER.
In the limit of large t-quark mass, the leading terms are given by~\cite{Akhundov:1985fc}:
\begin{eqnarray}
  \frac{\Delta_b(m_t^2)}{a_b}
 =
\frac{\alpha}{4 \pi \sin^2\theta_W}  |V_{tb}|^2 \frac{1}{2}
\left[ \frac{m_t^2}{M_W^2} + \left(\frac{8}{3}
+\frac{1}{6 \cos^2 \theta_W} \right) \log \frac{m_t^2}{M_W^2} \right],
\label{zbb}
\end{eqnarray}
where $V_{tb}$ is the $(t,b)$ Kobayashi-Maskawa mixing matrix element.
The ZFITTER calculations use the normalization $a_b=1$.
In the above, the matrix element 
\begin{eqnarray}
{\cal M}_b
&\sim&
\sqrt{ \frac{G_{\mu}}{\sqrt{2}} M_Z^2 }
\epsilon^{\beta} \sqrt{\rho_b^Z}
  a_b
 {\bar u} \left[ \gamma_{\beta} (1+\gamma_5) - 4 \sin^2\theta_W
 \kappa_b^Z
\gamma_{\beta}
 \right] u.
\label{zmate}
\end{eqnarray}
has been used.

The form factors for the $e^+e^-\to {\bar f} f$ scattering matrix element are related to those for the $Z\to {\bar f} f$ decay matrix 
element,
\begin{eqnarray}
\rho_{eb} &=&
\rho_{ed} -
  \frac{
  \Delta_b(m_t^2)
  }{a_b},
\label{rsbb}
\\
\kappa_b &=&
\kappa_d +
  \frac{\Delta_b(m_t^2)}{a_b},
\label{ksbb}
\\
\kappa_{eb} &=&
\kappa_{ed} +
  \frac{\Delta_b(m_t^2)}{a_b},
\label{kseb}
\end{eqnarray}
with $\kappa_e$ unchanged.

At the official ZFITTER webpage \href{http://sanc.jinr.ru/users/zfitter/}{http://sanc.jinr.ru/users/zfitter/} one may find the many 
additional publications on which the ZFITTER software is founded.

When the opening of LEP~1 at CERN approached,  with a potential to observe the decays $Z \to {\bar b}b$, several further one-loop 
calculations were published:
in 
October 1987 \cite{Bernabeu:1987me}, in January 1988 \cite{Beenakker:1988pv}, in January 1988 \cite{Jegerlehner:1988ak}, in July 1988 
\cite{Diakonos:1988zp}, in 1990 \cite{Lynn:1990hd}.
Just to mention, the one-loop terms of our present code for the $Z{\bar b}b$ vertex remained unpublished (A. Freitas).
\footnote{
The one-loop corrections to $Z \to {\bar b}b$ in the Gfitter package of 2007
(version of 15 June 2008 by J. Haller, A. Hoecker, M. Goebel \cite{gfitter-baak:2015}) are not
based on an independent calculation. That package makes use of the Standard Model implementation in ZFITTER \cite{Arbuzov:2005ma}.}

\subsection{Known higher order corrections to the $Z{\bar b}b$ vertex}
There are several higher-order corrections  to the $Z{\bar b}b$ vertex known:
\textcolor{black}{the $\OO(\alpha\as)$ QCD corrections} 
\cite{Djouadi:1987gn,Djouadi:1987di,Kniehl:1990yc,Kniehl:1991gu,Djouadi:1993ss,%
Czarnecki:1996ei,Harlander:1997zb}, 
\textcolor{black}{ partial corrections of order $\OO(\at\as^2)$ } \cite{Avdeev:1994db,Chetyrkin:1995ix},
\textcolor{black}{ partial corrections of order $\OO(\alpha^2\at)$ and $\OO(\at^3)$}  \cite{vanderBij:2000cg,Faisst:2003px}, 
 \textcolor{black}{the Standard Model two-loop prediction of $\MW$ from the Fermi constant $G_\mu$} \cite{Awramik:2003rn},
\textcolor{black}{partial corrections of order $\OO(\at\as^3)$ } \cite{Schroder:2005db,Chetyrkin:2006bj,Boughezal:2006xk},
\textcolor{black}{the fermionic electroweak two-loop corrections} \cite{Awramik:2008gi}.
Further references to be mentioned here are 
\cite{Bardin:1988xt,Awramik:2002wn,Onishchenko:2002ve,Freitas:2002ja,Awramik:2006uz,Freitas:2013dpa,Freitas:2014hra}.

\section{The bosonic $Z{\bar b}b$ topologies}
The bosonic electroweak two-loop corrections to the $Z{\bar b}b$ vertex were the last missing piece for the  prediction of 
$\seff{b}$.
We could use the calculational scheme as it was worked out in \cite{Awramik:2006uz,Awramik:2008gi} and work quoted therein.
We calculated the ${\cal O}(700)$ unknown bosonic Feynman diagrams with two methods, in order to have two independent numerical results. 
One method relied on sector decomposition, and we used the publicly available packages FIESTA 3 \cite{Smirnov:2013eza} and SecDec 3 
\cite{Borowka:2015mxa}. Both of them can apply contour deformation and are applicable to Minkowskian kinematics, as it is met here.
The second method uses Mellin-Barnes representations for the Feynman integrals. For this, one may use the MB suite, publicly available  at 
the MBtools webpage \cite{mbtools}, and software from the Katowice webpage \cite{Katowice-CAS:2007}. 
We had to develop two new tools. For the treatment of non-planar Feynman integrals, we developed AMBRE~3 
\cite{Bielas:2013v11,Dubovyk:2015yba,Dubovyk:201509v30x}, to complete the AMBRE versions 1 and 2 for planar cases \cite{Gluza:2007rt}. 
The package MBnumerics \cite{Usovitsch:201606} delivers a stable 8-digit numerical treatment of Feynman integrals with presently up to 
four dimensionless scales in the Minkowskian region.
It was also essential, that both methods can automatically treat ultraviolet and infrared singularities.
While the sector decomposition method had problems with few infrared divergent one- or two-scale integrals, the MB-method tends to fail 
for integrals with a larger number of scales. Nevertheless, we derived two precise, independent results for all the integrals needed.
Details of the complete calculation have been reported in \cite{Dubovyk:2016aqv,Dubovyk:2016ocz} and in the 
transparencies of this talk at LL2016 \cite{Riemann:ll2016}, so we may restrict ourselves here to some pedagogical remarks. 

\subsection{The non-planar one-scale integral $I_{15}(\textrm{0H0W0txZ})$}
\begin{figure}[t]
\begin{center}
 \includegraphics[width=5cm]{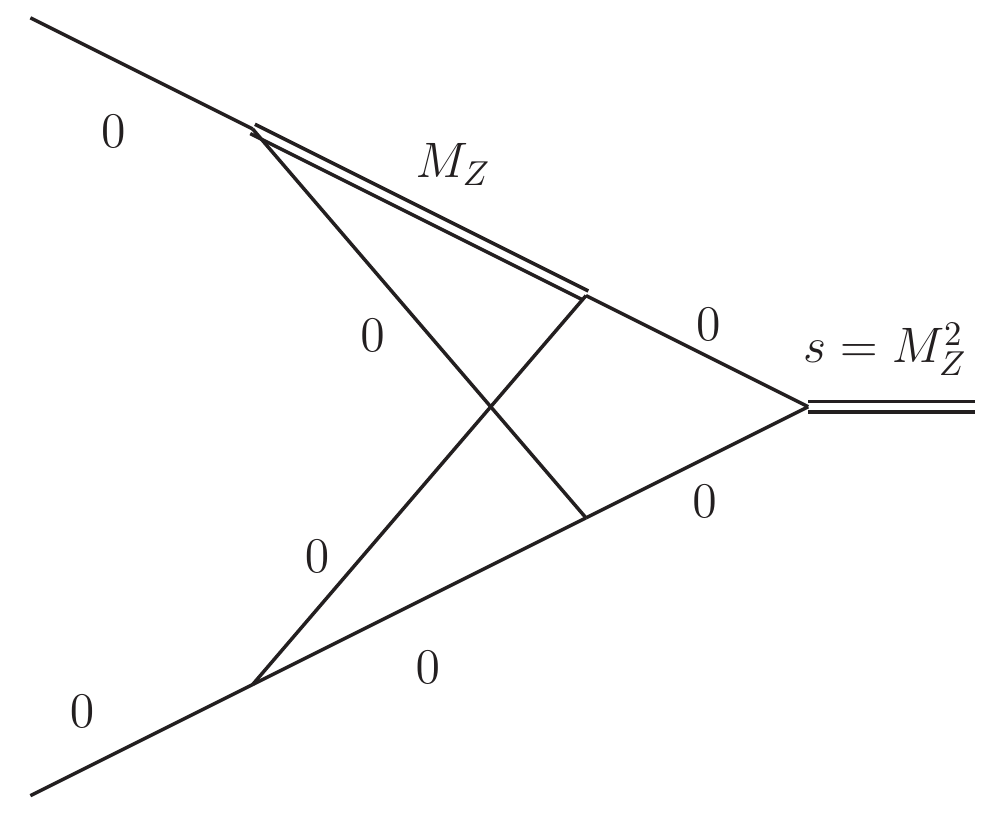}
\end{center}
\caption{\label{fig-fleischer}A non-planar vertex with one massive line.}
\end{figure}

As an example we compare several calculations of a non-planar two-loop vertex  integral  with one massive line and only one scale, 
$s=M_Z^2$, $I_{15}$(\textrm{0H0W0txZ}). It is shown in figure \ref{fig-fleischer} and depends on one parameter $s/M_Z^2= 1 + i \epsilon$.
A first calculation goes back to 1998 \cite{Fleischer:1998nb}, so we could use the result as a cross-check of our own calculation with the 
Mellin-Barnes method \cite{Dubovyk:2016ocz,Dubovyk:2016aqv}.
This was an important check, because an attempt to calculate the integral with the sector decomposition method in Minkowskian space-time 
failed.

The integral $I_{15}$(\textrm{0H0W0txZ}) is (up to some sign convention) the integral $N_3$ of 
\cite{Fleischer:1998nb}:
{
\begin{eqnarray}
- N_3 &=&  I_{15}(\textrm{0H0W0txZ})
\\ \nonumber
&=&
\frac{e^{2\gamma_E \epsilon}}{\pi^{d}}
\int \frac{d^dk_1d^dk_2}{
D[k_1, 0]  D[k_1-k_2, 0]D[k_2, 0] D[k_2 + p_2, 0] D[k_1 + k_2 + p_2, 0] D[k_1 - k_2 + p_1, M_Z]
}
.
\end{eqnarray}}

The MB-representation is derived 
with calls to the packages {PlanarityTest} \cite{Bielas:2013v11,Bielas:2013rja} and {AMBRE 3} 
\cite{%
Dubovyk:2015yba,Dubovyk:201509v30x}.
The $U$- and $F$-polynomials are:
\begin{alltt}{\footnotesize
Upoly1 =   x[1]x[2] + x[1]x[3] + x[2]x[3] + x[1]x[4] + x[3]x[4] + x[1]x[5] 
         + x[2]x[5] + x[4]x[5] + x[2]x[6] + x[3]x[6] + x[4]x[6] + x[5]x[6]

Fpoly2 = Upoly1 MZ^2 x[4] - s x[1]x[4]x[5] - s x[1]x[2]x[6] - s x[1]x[3]x[6]
                          - s x[2]x[3]x[6] - s x[1]x[4]x[6] - s x[1]x[5]x[6]
}\end{alltt}
A naive MB-representation would become high-dimensional, and it would be plagued by the occurrence of 
terms containing the ill-defined expression $\Gamma[0]$.
A dedicated introduction of Cheng-Wu variables leads to the following integrands for the $x$-integrations:
\begin{alltt}{
Upoly2 = v[1] + v[2] v[3]

Fpoly2 = + MZ^2 Upoly2 C[2]v[3] - s A[1]A[2]v[1]^2
         - s A[2]B[1]C[1]v[1]v[2]v[3] - s A[1]B[2]C[2]v[1]v[2]v[3]
}
\end{alltt}
The $x$-integrations over {\tt{v[i]}} can be easily performed, and we remain, from the four additive terms in {\tt{Fpoly2}}, 
with a three-dimensional MB-integral:

\begin{alltt}{\footnotesize
N3 ~  (-s)^(-2-2eps)  Gamma[-eps]
      ({-(s/MZ^2)})^-z2 Gamma[-eps-z1] Gamma[-z1] Gamma[-eps-z2] Gamma[-z2] 
        Gamma[-1-2eps-z1-z3] Gamma[-1-2eps-z2-z3] Gamma[-1-2eps-z1-z2-z3] 
        Gamma[-z3] Gamma[1+z3]^2 Gamma[1+z1+z3] Gamma[2+2eps+z1+z2+z3])
        / (Gamma[-2eps-z1] Gamma[-3eps-z2] Gamma[-2eps-z2] Gamma[-2eps-z1-z2])
}\end{alltt}

When continuing  in \texttt{eps} with the package {MB.m} \cite{Czakon:2005rk}, we derive for vanishing, but finite $\epsilon$ a 
two-dimensional and a three-dimensional MB-representation:

\begin{alltt}{\footnotesize
N3 ~
\{
MBint[ ((-s)^(-2-2eps) Gamma[-2eps] Gamma[-eps]
({-(s/MZ^2)})^-z2 Gamma[-eps-z2] Gamma[-z2]^2 Gamma[1+z2] Gamma[-1-2eps-z2-z3] 
Gamma[-z3] Gamma[1+z3] Gamma[1+eps+z3] Gamma[1+2eps+z3])
 / (Gamma[-3eps-z2] Gamma[-2eps-z2] Gamma[1-z2+z3]), 
\{\{eps->0\},\{{z2}->-0.42644,{z3}->-0.826119\}\} ],

MBint[ ((-s)^(-2-2eps) Gamma[-eps]
 ({-(s/MZ^2)})^-z2 Gamma[-eps-z1] Gamma[-z1] Gamma[-eps-z2] Gamma[-z2] 
 Gamma[-1-2eps-z1-z3] Gamma[-1-2eps-z2-z3] Gamma[-1-2eps-z1-z2-z3] 
 Gamma[-z3]  Gamma[1+z3]^2 Gamma[1+z1+z3]  Gamma[2+2eps+z1+z2+z3]) 
 / (Gamma[-2eps-z1] Gamma[-3eps-z2] Gamma[-2eps-z2] Gamma[-2 eps-z1-z2]), 
\{\{eps->0\},\{{z1}->-0.268281,{z2}->-1.00065,{z3}->-0.171895\}\} ]
\}
}\end{alltt}

Applying MB \cite{Czakon:2005rk} and MBnumerics \cite{Usovitsch:201606}, we get: 
\begin{eqnarray}
\label{eq-n3ambre}
 N_3 
&=& 
\frac{1}{\epsilon^2} (1.23370055013617 - i~6.20475892887384 \times 10^{-13} )
\\\nonumber
&&
+~\frac{1}{\epsilon}
( 2.8902545096591976 + i~3.875784585038738 )
\\\nonumber
&&
+~ (\textcolor{black}{-0.778599608}3247692  - i~\textcolor{black}{4.1235126}00516016 )
.
\end{eqnarray}

The integral N3, according to equation (D.11) of  \cite{Fleischer:1998nb}, is with $z~=~s/M_Z^2~=~1+i\epsilon$:
{
\begin{eqnarray}\label{eq-n3}
N_3 &=&
\frac{1}{s^2} 
\sum_{n=1}^{\infty} \textcolor{black}{(-z)^n}
\Biggl\{ 
\frac{1}{\epsilon^2}
        \Biggl[-\frac{1}{2} \zeta_2 + K_2( n - 1)
\Biggr] 
\\ \nonumber
&&+~
 \frac{1}{\epsilon}  
\Bigl[-\frac{1}{2} \zeta_3 - 2\zeta_2 S_1(n-1) + 
          2 S_3(n-1) - 2 K_3( n - 1) + 
          4 S_1(n-1) K_2( n - 1) 
\\ \nonumber
&&
+ ~ (\zeta_2 - S_2(n - 1)
) 
           \textcolor{black}{\ln(-z)}
\Biggr] 
\\ \nonumber 
&& + ~
\Biggl[-\zeta_4 - 2 \zeta_3 S_1(n-1) - 
          7 \zeta_2 S_2(n - 1) - 4 \zeta_2 S_1(n-1)^2 + 
          7 \zeta_2 K_2( n - 1) - \frac{7}{2} S_4( n - 1)
\\ \nonumber 
&& + ~  \frac{7}{2} S_2(n - 1)^2 + 6 S_1(n-1) S_3(n-1) + 
          2 S_{13}( n - 1) + 8 K_4( n - 1) 
\\ \nonumber 
&& -~
          8 S_1(n-1) K_3( n - 1) 
+ 
          8 S_1(n-1)^2 
           K_2( n - 1) 
\\ \nonumber 
&& 
+  ~
\Bigl(\zeta_3 + 4 \zeta_2 S_1(n-1) - 
             S_{12}( n - 1) 
- 3 S_1(n-1) S_2(n - 1) - 
             4 K_3( n - 1)  
\Bigr) \textcolor{black}{\ln(-z) }
\\ \nonumber 
&& 
+ ~
\Bigl(
-\zeta_2 + \frac{1}{2} S_2(n - 1) + K_2( n - 1) \Bigr)\textcolor{black}{\ln^2(-z)}
\Biggr]
\Biggr\}
.
\end{eqnarray}
}

The expression contains harmonic sums \cite{Kazakov:1986mu,Kotikov:1987mw,Kazakov:1992xj}:
\begin{eqnarray}
S_a(n)&=&\sum_{j=1}^{n} {1}/{j^a},
\\
K_a(n) &=& - \sum_{j=1}^{n} (-1)^{j}/{j^a},
\\
S_{ab}(n) &=& \sum_{j=1}^{n} {S_b(j-1)}/{j^a}, 
\end{eqnarray}

The sum $N_3$ converges both in the Euklidean and the Minkowskian kinematics, but very slowly, so that it would need many terms in order to 
get our accuracy goal of eight digits.
The {N3 evaluated with 200 terms gives e.g.: }
\begin{verbatim}
time = 4.060519 sec       for 200 terms of the sum

N3 = (0.4 + 4 x I) 

   + 1/eps x (2.8 + 3.87 x I) + 1 /eps^2 x (1.23 + 0 x I)
                                 
\end{verbatim}
The agreement, with several thousand terms (few hours running time), is suffiently good in order to see that the results from the 
numerical 
MB-approach 
are reasonable. 
One may improve the comparison.
 In fact, in appendix~E of   \cite{Fleischer:1998nb}, the necessary  harmonic sums are explicitly performed.
We derive:\footnote{In \cite{Fleischer:1998nb}, the overall sign of (E.7) is wrong, and in the r.h.s. of (E.36) one has to replace 
$S_{1,3}$ under the integral by $S_{1,2}$ and to change the sign of $2\ln(1-z)$. We thank A. Kotikov for  clarifying this.}

\begin{eqnarray}
\phi(z) = \frac{1}{2} S_{1, 2}(z^2) - S_{1, 2}(z) - 
  S_{1, 2}(-z) + \ln(1 - z)\text{Li}_2(-z)
\end{eqnarray}
and
\begin{eqnarray}
-\frac{s^2z}{1+z} ~{\color{black}{ N_3}} &=&
~~ {\color{black}{\frac{1}{\epsilon^2}}} \left[- \frac{\zeta_2}{2}-\text{Li}_2(z) \right]
\\ \nonumber
&&
+ ~{\color{black}{\frac{1}{\epsilon}}} \biggl[
-\frac{1}{2}\zeta_3 -2 \zeta_2 \text{Li}_2(-z) 
+2 \text{Li}_3(-z) 
+ 2 \text{Li}_3(z) 
+4 \bigl[\phi(-z) -S_{1,2}(z)-\text{Li}_3(-z)\bigr]
\\ \nonumber
&&
+ ~(\zeta_2- \text{Li}_2(-z) 
)\ln(-z)
\biggr]
\\ \nonumber
&& 
-~\frac{s^2z}{1+z}~ {\color{black}{N_3^{\text{const}} }}
\end{eqnarray}

Further,
\begin{eqnarray}
\frac{s^2(-z)}{1+z}~ {\color{black}{N_3^{\text{const}}}} &=&
-\zeta_4 
    + 2 \zeta_3 \ln(1+z)
- 7 \zeta_2 \text{Li}_2(-z) 
- 4~  \zeta_2  \left[ -\ln(1-r)^2 + \text{Li}_2(-z) \right]
\\ \nonumber
&&
 + 
 7~ \zeta_2  \text{Li}_2(z) 
 - \frac{7}{2}  \text{Li}_4(-z)
 +  \frac{7}{2} \left[\text{Li}_2(-z)^2 + \text{Li}_4(-z)-S_{2,2}(-z) \right] 
\\ \nonumber
&&+~ 6 \bigl[ (\text{Li}_4(-z) - \ln(1+z) \text{Li}_3(-z)-\frac{1}{2}\text{Li}_2(-z)^2+S_{2,2}(-z) \bigr]
\\ \nonumber
&&
+ 
 2 ~  \left[-\ln(1+z) \text{Li}_3(-z)-\frac{1}{2}\text{Li}_2(-z)^2 \right]
+ 8  \text{Li}_4(z) 
\\ \nonumber 
&& -~ 8   \left[-\text{Li}_4(z)-S_{2,2}(z) - \int_0^{-z} \frac{dt}{1-t} \text{Li}_3(-t)\right]
\\ \nonumber
&&
+ 8 
\biggl[
-\text{Li}_4(z)-\frac{1}{2}S_{2,2}(z) - 2S_{1,3}(z)+\phi(-z)[\ln(-z)-2\ln(1+z)]
\\ \nonumber
&&
- \int_0^{-z} \frac{dt}{1-t} [ ( 2\ln(1-t)-\ln(t) )\text{Li}_2(-t) +2\text{Li}_3(-t)+2S_{1,2}(-t)]
\biggr]
\\ \nonumber
&&
+ 
\Bigl[\zeta_3 - 4 \zeta_2 \ln(1+z) + \bigl[\ln(1+z)\text{Li}_2(-z)+S_{1,2}(-z)\bigr] 
\\ \nonumber
&&-~ 3  \bigl(\text{Li}_3(-z)+(-\ln(1+z)\text{Li}_2(-z)-S_{1,2}(-z)\bigr)
 -  4  \text{Li}_3(z)  
\Bigr] 
\ln(-z)
\\ \nonumber
&&
 +
\Bigl[
-\zeta_2 
+ \frac{1}{2} \text{Li}_2(-z)
 + \text{Li}_2(z) \Biggr]
\ln^2(-z)
\end{eqnarray}
The result is expressed in terms of polylogarithms, plus the few harmonic polylogarithms which are needed to close the basis of weight 
four.
 For a systematic numerical calculation of the expressions here in terms of  harmonic polylogarithms $H[x,a,b,c,d], a,b,c,d = \pm 1 , 
0$, see e.g.  appendix~B of \cite{Czakon:2004wm}, which is implemented in the Mathematica 
package {HPL4num.m}  \cite{Riemann:2004zz}
and checked with the Mathematica package HPL \cite{Maitre:2005uu}.
The most compact representation of the integral at the $Z$ boson mass shell was obtained with the aid of Jacob Ablinger, 
Johannes Bl\"umlein, Carsten Schneider and Arnd Behring (priv. commun.):
\begin{eqnarray}
N_{3,-2}(z) 
&=& (z \zeta_2)/(2 (1 + z)) + (z \text{Li}_2(z))/(1 + z),
\\
N_{3,-2}(1+i~\epsilon) &=& \frac{3 \zeta_2}{4} ,
\\
N_{3,-1}(z) &=& 
\frac{z}{2 (z+1)}
\bigl[ 
(-4 H(-1,z) (2 H(0,1,z)+\zeta_2)
+8 H(0,-1,1,z)
\\ \nonumber
&&
-~4 H(0,0,-1,-z)
+4 H(0,0,-1,z)+8 H(0,1,-1,z)+8 H(0,1,1,z)
\\ \nonumber
&&-~2 \ln (-z) H(0,-1,z)
-2~\zeta_2 \ln (-z)+\zeta_3)\bigr] ,
\\
N_{3,-1}(1+i~\epsilon) &=& 
-3 \ln(2) \zeta_2 + \frac{21}{4} \zeta_3 + i~\frac{3}{4}  \pi \zeta_2 .
\end{eqnarray}
We confirm with $N_{3}(1+i~\epsilon)$ the \textcolor{black}{9 digits accuracy} obtained with AMBRE/MB/MBnumerics given in 
\eqref{eq-n3ambre}:
\begin{eqnarray}
N_{3,0}(1+i~\epsilon) &=&
24 \text{Li}_4(1/2) + \ln^4(2) - \frac{351 }{80} \zeta^2_2
- I~3 \ln(2)2 \pi \zeta_2  - i~\frac{63}{16} \pi \zeta_3 
\\\nonumber
&=&
0.77859960898762168563452805690 
\\\nonumber
&&
- i~ 4.12351259333642272648103365383 
.
\end{eqnarray}

At the end of this section we like to mention, as an additional calculational alternative, a quite recent numerical approach to 
single-scale Feynman integrals \cite{Gluza:2016fwh}.


\section{Results}
The electroweak bosonic two-loop contribution to the weak mixing angle is:
\begin{eqnarray}
 \Delta\kappa_\Pb^{(\alpha^2,\rm bos)} = - 0.9855 \times 10^{-4}.
\end{eqnarray}
The value  $-1.0276 \times 10^{-4}$, presented as preliminary result at LL2016, was based on the input parameter list of 
\cite{Riemann:ll2016} which differs slightly from the input list of table \ref{tab:input} of \cite{Dubovyk:2016aqv}, which is applied here.
This value amounts to about $\frac{1}{4}$ of the leptonic corrections to $\kappa_b$ and $\seff{b}$.
The corrections to the weak mixing angle are shown in table \ref{tab:orders}.
The biggest corrections come from the one-loop electroweak contributions, followed by mixed electroweak-QCD 
corrections of order $\alpha\alpha_s$.
All the other corrections, including the new bosonic electroweak two-loop corections,  are of the same order, at the $10^{-4}$ level. For a 
per mille measurement, it is good to know them, but they will not influence the data analysis numerically.

\begin{table}[t]
\caption{Reference values used in the numerical analysis, from
Ref.~\cite{Agashe:2014kda}.}
\label{tab:input}
\renewcommand{\arraystretch}{1.25}
%
\begin{center}
\begin{tabular}{lll}
\hline
Parameter & Value & Range \\
\hline
$M_{\mathrm{Z}}$ & 91.1876 GeV & $\pm0.0042~\mbox{GeV}$ \\
$\Gamma_\mathrm{Z}$ & 2.4952 GeV & \\
$M_{\mathrm{W}}$ & 80.385 GeV & $\pm0.030~\mbox{GeV}$ \\
$\Gamma_\mathrm{W}$ & 2.085 GeV & \\
$M_{\mathrm{H}}$ & 125.1 GeV & $\pm5.0~\mbox{GeV}$ \\
$m_{\mathrm{t}}$ & 173.2 GeV & $\pm4.0~\mbox{GeV}$ \\
$\alpha_{\mathrm{s}}$ & 0.1184 & $\pm0.0050$ \\
$\Delta\alpha$ & $0.0590$ & $\pm0.0005$ \\
\hline
\end{tabular}
\end{center}
\end{table}

\normalsize

\begin{table}[t]
\caption[]{%
Comparison of different orders of radiative corrections to
$\Delta\kappa_\mathrm{b}$, using the input parameters in \ref{tab:input}. Numerical values taken from \cite{Dubovyk:2016aqv}.
}
\label{tab:orders}
\renewcommand{\arraystretch}{1.3}
\begin{center}
\begin{tabular}{ll}
\hline
Order & Value [$10^{-4}$] \\
\hline
$\alpha$ & 468.945 \\
$\alpha\alpha_{\mathrm{s}}$ & $-42.655$ \\
$\alpha_{\mathrm{t}}\alpha_{\mathrm{s}}^2$ & $-7.074$ \\
$\alpha_{\mathrm{t}}\alpha_{\mathrm{s}}^3$ & $-1.196$ \\
\hline
%
\hfill
$\alpha_{\mathrm{t}}^2\alpha_{\mathrm{s}}$ & 1.362 \\
$\alpha_{\mathrm{t}}^3$ & 0.123 \\
$\alpha^2_{\mathrm{ferm}}$ & 3.866 \\
$\alpha^2_{\mathrm{bos}}$ & $-0.986$ \\
\hline
\end{tabular}
\end{center}
\end{table}

For the corresponding fitting formula for $\seff{b}$, we refer to \cite{Dubovyk:2016aqv}.
An analysis tool for the consistent 1 per mille treatment of realistic observables, pseudo-observables, and two-loop predictions to them, 
is not available for the $Z$ boson resonance, although ZFITTER is a very good approximation and suffices for the presently available 
accuracy of data.

\section*{Acknowledgements}
We would like to thank Peter Marquard for discussions and Peter Uwer
and his group ``Phenomenology of Elementary
Particle Physics beyond the Standard Model'' at Humboldt-Universit\"at
zu Berlin for providing computer resources.

The work of \textit{I.D.}\ is supported by a research grant of
Deutscher Akademischer Austausch\-dienst DAAD
and by Deutsches
Elektronensychrotron DESY.
The work of \textit{J.G.}\ is supported by the Polish National Science Centre NCN,
Grant No.
{DEC-2013/11/B/ST2/04023}.
The work of \textit{A.F.}\ is supported in part by the U.S.\
{National Science Foundation} under grant PHY-1519175.
{The work of \textit{T.R.}\ is supported in part by an
{Alexander von Humboldt Polish Honorary Research Fellowship}.
The work of \textit{J.U.}\ is supported by Graduiertenkolleg 1504
``Masse, Spektrum, Symmetrie'' of
{Deutsche Forschungsgemeinschaft} (DFG).}
  \textit{A.F.} gratefully acknowledges the hospitality of the Kavli
Institute for
Theoretical Physics China during the final stages of this project.


\end{document}